
%
\def\unredoffs{}
\tolerance=1000\hfuzz=2pt
\catcode`\@=11 
\ifx\hyperdef\UNd@FiNeD\def\hyperdef#1#2#3#4{#4}\def\hyperref#1#2#3#4{#4}\def\href#1#2{#2}\fi
\magnification=1200\unredoffs\baselineskip=16pt plus 2pt minus 1pt
\def\Date#1{\vfill\leftline{#1}\tenpoint\supereject%
\footline={\hss\tenrm\hyperdef\hypernoname{page}\folio\folio\hss}}%

{\count255=\time\divide\count255 by 60 \xdef\hourmin{\number\count255}
 \multiply\count255 by-60\advance\count255 by\time
 \xdef\hourmin{\hourmin:\ifnum\count255<10 0\fi\the\count255}
}
\def\date{\number\day.\number\month.\number\year\ at \hourmin}


\def\nolabels{\def\wrlabeL##1{}\def\eqlabeL##1{}\def\reflabeL##1{}}
\def\writelabels{\def\wrlabeL##1{\leavevmode\vadjust{\rlap{\smash%
{\line{{\escapechar=` \hfill\rlap{\sevenrm\hskip.03in\string##1}}}}}}}%
\def\eqlabeL##1{{\escapechar-1\rlap{\sevenrm\hskip.05in\string##1}}}%
\def\reflabeL##1{\noexpand\llap{\noexpand\sevenrm\string\string\string##1}}}
\nolabels

\global\newcount\secno \global\secno=0
\global\newcount\meqno \global\meqno=1
\def\s@csym{}

\def\newsec#1\par{\global\advance\secno by1%
{\toks0{#1}\message{(\the\secno. \the\toks0)}}%
\global\subsecno=0\eqnres@t\let\s@csym\secsym\xdef\secn@m{\the\secno}\noindent
{\bf\hyperdef\hypernoname{section}{\the\secno}{\the\secno.} #1}%
\writetoca{{\string\hyperref{}{section}{\the\secno}{\bf \the\secno\quad}} {\bf #1}}\par%
\nobreak\medskip\nobreak\noindent\ignorespaces}
\def\eqnres@t{\xdef\secsym{\the\secno.}\global\meqno=1\bigbreak\bigskip}
\def\sequentialequations{\def\eqnres@t{\bigbreak}}\xdef\secsym{}

\global\newcount\subsecno \global\subsecno=0
\def\subsec#1\par{\global\advance\subsecno by1%
{\toks0{#1}\message{(\s@csym\the\subsecno. \the\toks0)}}%
\global\subsubsecno=0%
\ifnum\lastpenalty>9000\else\bigbreak\fi
\noindent{\it\hyperdef\hypernoname{subsection}{\secn@m.\the\subsecno}%
{\secn@m.\the\subsecno.} #1}\writetoca{\string\hskip1.45cm
{\string\hyperref{}{subsection}{\secn@m.\the\subsecno}{\secn@m.\the\subsecno.}}
{#1}}\par\nobreak\medskip\nobreak\noindent\ignorespaces}

\def\appendix#1#2{\global\meqno=1\global\subsecno=0\xdef\secsym{\hbox{#1.}}%
\bigbreak\bigskip\noindent{\bf Appendix \hyperdef\hypernoname{appendix}{#1}%
{#1.} #2}{\toks0{(#1. #2)}\message{\the\toks0}}%
\xdef\s@csym{#1.}\xdef\secn@m{#1}%
\writetoca{{\string\hyperref{}{appendix}{#1}{\bf {#1}\quad}} {\bf #2}}%
\par\nobreak\medskip\nobreak}

%
\def\checkm@de#1#2{\ifmmode{\def\f@rst##1{##1}\hyperdef\hypernoname{equation}%
{#1}{#2}}\else\hyperref{}{equation}{#1}{#2}\fi}
\def\eqnn#1{\DefWarn#1\xdef #1{(\noexpand\relax\noexpand\checkm@de%
{\s@csym\the\meqno}{\secsym\the\meqno})}%
\wrlabeL#1\writedef{#1\leftbracket#1}\global\advance\meqno by1}
\def\f@rst#1{\c@t#1a\em@ark}\def\c@t#1#2\em@ark{#1}
\def\eqna#1{\DefWarn#1\wrlabeL{#1$\{\}$}%
\xdef #1##1{(\noexpand\relax\noexpand\checkm@de%
{\s@csym\the\meqno\noexpand\f@rst{##1}1}{\hbox{$\secsym\the\meqno##1$}})}
\writedef{#1\numbersign1\leftbracket#1{\numbersign1}}\global\advance\meqno by1}
\def\eqn#1#2{\DefWarn#1%
\xdef #1{(\noexpand\hyperref{}{equation}{\s@csym\the\meqno}%
{\secsym\the\meqno})}$$#2\eqno(\hyperdef\hypernoname{equation}%
{\s@csym\the\meqno}{\secsym\the\meqno})\eqlabeL#1$$%
\writedef{#1\leftbracket#1}\global\advance\meqno by1}
\def\xeqn{\expandafter\xe@n}\def\xe@n(#1){#1}
\def\xeqna#1{\expandafter\xe@n#1}
\def\eqns#1{(\e@ns #1{\hbox{}})}
\def\e@ns#1{\ifx\UNd@FiNeD#1\message{eqnlabel \string#1 is undefined.}%
\xdef#1{(?.?)}\fi{\let\hyperref=\relax\xdef\next{#1}}%
\ifx\next\em@rk\def\next{}\else%
\ifx\next#1\xeqn#1\else\def\n@xt{#1}\ifx\n@xt\next#1\else\xeqna#1\fi
\fi\let\next=\e@ns\fi\next}

\def\DefWarn#1{\ifx\UNd@FiNeD#1\else
\immediate\write16{*** WARNING: the label \string#1 is already defined ***}\fi}
%
\newskip\footskip\footskip14pt plus 1pt minus 1pt 
\def\footnotefont{\ninepoint}\def\f@t#1{\footnotefont #1\@foot}
\def\f@@t{\baselineskip\footskip\bgroup\footnotefont\aftergroup\@foot\let\next}
\setbox\strutbox=\hbox{\vrule height9.5pt depth4.5pt width0pt}
\global\newcount\ftno \global\ftno=0
\def\foot{\global\advance\ftno by1\def\foot@rg{\hyperref{}{footnote}%
{\the\ftno}{\the\ftno}\xdef\foot@rg{\noexpand\hyperdef\noexpand\hypernoname%
{footnote}{\the\ftno}{\the\ftno}}}\footnote{$^{\foot@rg}$}}
%
%
%
\global\newcount\refno \global\refno=1
\newwrite\rfile
\def\ref{[\hyperref{}{reference}{\the\refno}{\the\refno}]\nref}
\def\nref#1{\DefWarn#1%
\xdef#1{[\noexpand\hyperref{}{reference}{\the\refno}{\the\refno}]}%
\writedef{#1\leftbracket#1}%
\ifnum\refno=1\immediate\openout\rfile=\jobname.refs\fi
\chardef\wfile=\rfile\immediate\write\rfile{\noexpand\item{[\noexpand\hyperdef%
\noexpand\hypernoname{reference}{\the\refno}{\the\refno}]\ }%
\reflabeL{#1\hskip.31in}\pctsign}\global\advance\refno by1\findarg}
\def\findarg#1#{\begingroup\obeylines\newlinechar=`\^^M\pass@rg}
{\obeylines\gdef\pass@rg#1{\writ@line\relax #1^^M\hbox{}^^M}%
\gdef\writ@line#1^^M{\expandafter\toks0\expandafter{\striprel@x #1}%
\edef\next{\the\toks0}\ifx\next\em@rk\let\next=\endgroup\else\ifx\next\empty%
\else\immediate\write\wfile{\the\toks0}\fi\let\next=\writ@line\fi\next\relax}}
\def\striprel@x#1{} \def\em@rk{\hbox{}}
\def\lref{\begingroup\obeylines\lr@f}
\def\lr@f#1#2{\DefWarn#1\gdef#1{\let#1=\UNd@FiNeD\ref#1{#2}}\endgroup\unskip}

\def\addref#1{\immediate\write\rfile{\noexpand\item{}#1}} 
\def\listrefs{\vfill\supereject\immediate\closeout\rfile\writestoppt
\baselineskip=\footskip\centerline{{\bf References}}\bigskip{\parindent=20pt%
\frenchspacing\escapechar=` \input \jobname.refs\vfill\eject}\nonfrenchspacing}
\def\startrefs#1{\immediate\openout\rfile=\jobname.refs\refno=#1}
\def\xref{\expandafter\xr@f}\def\xr@f[#1]{#1}
\def\refs#1{\count255=1[\r@fs #1{\hbox{}}]}
\def\r@fs#1{\ifx\UNd@FiNeD#1\message{reflabel \string#1 is undefined.}%
\nref#1{need to supply reference \string#1.}\fi%
\vphantom{\hphantom{#1}}{\let\hyperref=\relax\xdef\next{#1}}%
\ifx\next\em@rk\def\next{}%
\else\ifx\next#1\ifodd\count255\relax\xref#1\count255=0\fi%
\else#1\count255=1\fi\let\next=\r@fs\fi\next}
%

%
\newwrite\ffile\global\newcount\figno \global\figno=1
\def\fig{fig.~\hyperref{}{figure}{\the\figno}{\the\figno}\nfig}
\def\nfig#1{\DefWarn#1%
\xdef#1{fig.~\noexpand\hyperref{}{figure}{\the\figno}{\the\figno}}%
\writedef{#1\leftbracket fig.\noexpand~\xfig#1}%
\ifnum\figno=1\immediate\openout\ffile=\jobname.figs\fi\chardef\wfile=\ffile%
{\let\hyperref=\relax
\immediate\write\ffile{\noexpand\medskip\noexpand\item{Fig.\ %
\noexpand\hyperdef\noexpand\hypernoname{figure}{\the\figno}{\the\figno}. }
\reflabeL{#1\hskip.55in}\pctsign}}\global\advance\figno by1\findarg}
\def\xfig{\expandafter\xf@g}\def\xf@g fig.\penalty\@M\ {}
\def\figs#1{figs.~\f@gs #1{\hbox{}}}
\def\f@gs#1{{\let\hyperref=\relax\xdef\next{#1}}\ifx\next\em@rk\def\next{}\else
\ifx\next#1\xfig #1\else#1\fi\let\next=\f@gs\fi\next}
%
\def\figin{\epsfcheck\figin}\def\figins{\epsfcheck\figins}
\def\epsfcheck{\ifx\epsfbox\UnDeFiNeD
\message{(NO epsf.tex, FIGURES WILL BE IGNORED)}
\gdef\figin##1{\vskip2in}\gdef\figins##1{\hskip.5in}
\else\message{(FIGURES WILL BE INCLUDED)}%
\gdef\figin##1{##1}\gdef\figins##1{##1}\fi}
\def\DefWarn#1{}
\def\figinsert{\goodbreak\topinsert}
\def\ifig#1#2#3{\DefWarn#1\xdef#1{fig.~\the\figno}
\writedef{#1\leftbracket fig.\noexpand~\the\figno}%
\figinsert\figin{\centerline{#3}}
\smallskip
\leftskip=20pt \rightskip=20pt
\baselineskip12pt\noindent
{{\bf Fig.~\the\figno}\ \ninepoint #2}
\medskip
\global\advance\figno by1\par\endinsert}
\newwrite\lfile
{\escapechar-1\xdef\pctsign{\string\%}\xdef\leftbracket{\string\{}
\xdef\rightbracket{\string\}}\xdef\numbersign{\string\#}}
\def\writedefs{\immediate\openout\lfile=label.defs \def\writedef##1{%
{\let\hyperref=\relax\let\hyperdef=\relax\let\hypernoname=\relax
 \immediate\write\lfile{\string\def\string##1\rightbracket}}}}%
\def\writestop{\def\writestoppt{\immediate\write\lfile{\string\pageno
 \the\pageno\string\startrefs\leftbracket\the\refno\rightbracket
 \string\def\string\secsym\leftbracket\secsym\rightbracket
 \string\secno\the\secno\string\meqno\the\meqno}\immediate\closeout\lfile}}
\def\writestoppt{}\def\writedef#1{}

\def\seclab#1{\DefWarn#1%
\xdef #1{\noexpand\hyperref{}{section}{\the\secno}{\the\secno}}%
\writedef{#1\leftbracket#1}\wrlabeL{#1=#1}}
\def\subseclab#1{\DefWarn#1%
\xdef #1{\noexpand\hyperref{}{subsection}{\the\secno.\the\subsecno}%
{\the\secno.\the\subsecno}}\writedef{#1\leftbracket#1}\wrlabeL{#1=#1}}
\def\applab#1{\DefWarn#1%
\xdef #1{\noexpand\hyperref{}{appendix}{\secn@m}{\secn@m}}%
\writedef{#1\leftbracket#1}\wrlabeL{#1=#1}}
\newwrite\tfile \def\writetoca#1{}
\def\leaderfill{\leaders\hbox to 1em{\hss.\hss}\hfill}
\def\writetoc{\immediate\openout\tfile=\jobname.toc
   \def\writetoca##1{{\edef\next{\write\tfile{\noindent ##1
   \string\leaderfill{
   \string\hyperref{}{page}{\noexpand\number\pageno}%
   {\noexpand\number\pageno}} \par}}\next}}
}
\newread\ch@ckfile
\def\listtoc{\immediate\closeout\tfile\immediate\openin\ch@ckfile=\jobname.toc
\ifeof\ch@ckfile\message{no file \jobname.toc, no table of contents this pass}%
\else\closein\ch@ckfile\centerline{\bf Contents}\nobreak\medskip%
{\baselineskip=16pt\footnotefont\parskip=0pt\catcode`\@=11\input\jobname.toc
\catcode`\@=12\bigbreak\bigskip}\fi}
\catcode`\@=12 
\def\tenpoint{\def\rm{\fam0\tenrm}
\textfont0=\tenrm \scriptfont0=\sevenrm \scriptscriptfont0=\fiverm
\textfont1=\teni  \scriptfont1=\seveni  \scriptscriptfont1=\fivei
\textfont2=\tensy \scriptfont2=\sevensy \scriptscriptfont2=\fivesy
\textfont\itfam=\tenit \def\it{\fam\itfam\tenit}\def\footnotefont{\ninepoint}%
\textfont\bffam=\tenbf \def\bf{\fam\bffam\tenbf}\def\sl{\fam\slfam\tensl}\rm}
\font\ninerm=cmr9 \font\sixrm=cmr6 \font\ninei=cmmi9 \font\sixi=cmmi6
\font\ninesy=cmsy9 \font\sixsy=cmsy6 \font\ninebf=cmbx9
\font\nineit=cmti9 \font\ninesl=cmsl9 \skewchar\ninei='177
\skewchar\sixi='177 \skewchar\ninesy='60 \skewchar\sixsy='60
\def\ninepoint{\def\rm{\fam0\ninerm}
\textfont0=\ninerm \scriptfont0=\sixrm \scriptscriptfont0=\fiverm
\textfont1=\ninei \scriptfont1=\sixi \scriptscriptfont1=\fivei
\textfont2=\ninesy \scriptfont2=\sixsy \scriptscriptfont2=\fivesy
\textfont\itfam=\ninei \def\it{\fam\itfam\nineit}\def\sl{\fam\slfam\ninesl}%
\textfont\bffam=\ninebf \def\bf{\fam\bffam\ninebf}\rm}
%
\hyphenation{anom-aly anom-alies coun-ter-term coun-ter-terms}

\global\newcount\subsubsecno \global\subsubsecno=0
\def\subsubsec#1\par{\global\advance\subsubsecno by1%
{\toks0{#1}\message{(\the\secno\the\subsecno\the\subsubsecno. \the\toks0)}}%
\ifnum\lastpenalty>9000\else\bigbreak\fi
\noindent{\it\hyperdef\hypernoname{subsubsection}{\the\secno.\the\subsecno\the\subsubsecno}%
{\the\secno.\the\subsecno.\the\subsubsecno.} #1}
\par\nobreak\medskip\nobreak\noindent\ignorespaces}

\def\DefWarn#1{}
\def\tikzcaption#1#2{\DefWarn#1\xdef#1{Fig.~\the\figno}
\writedef{#1\leftbracket Fig.\noexpand~\the\figno}%
{
\smallskip
\leftskip=20pt \rightskip=20pt \baselineskip12pt\noindent
{{\bf Fig.~\the\figno}\ \ninepoint #2}
\bigskip
\global\advance\figno by1 \par}}

\def\ntoalpha#1{%
\ifcase#1%
@%
\or A\or B\or C\or D\or E\or F\or G\or H\or I
\fi
}

\global\newcount\appno \global\appno=1
\def\applab#1{\xdef #1{\ntoalpha\appno}\writedef{#1\leftbracket#1}\wrlabeL{#1=#1}
\global\advance\appno by1}

\def\preprint#1 #2\par{\rightline{\vbox{\baselineskip12pt\hbox{#1}\hbox{#2}}}\vskip2cm}
%
\def\title#1\par{\centerline{\bf #1}\nopagenumbers\pageno=0}
\def\author#1\par{\bigskip\bigskip\centerline{#1}}

\newcount\addressno

\def\email#1#2{\unskip$^#1$\footnote{\null}{\kern-\parindent \llap{$^#1$\hskip1pt}email: #2}}

\def\startcenter{%
  \par
  \begingroup
  \leftskip=0pt plus 1fil
  \rightskip=\leftskip
  \parindent=0pt
  \parfillskip=0pt
}
\def\stopcenter{\endgroup}

\def\address{\bigskip%
  \ifnum\the\addressno=0\else\stopcenter\endgroup\fi
  \advance\addressno by 1%
  \begingroup
  \startcenter
  \it
  \obeylines
  \addressAux
}
\def\addressAux#1{#1}

\def\abstract{\stopcenter\endgroup\bigskip\bigskip\noindent}

\def\Dsl{\,\raise.15ex\hbox{/}\mkern-13.5mu D} 
\def\dsl{\raise.15ex\hbox{/}\kern-.57em\partial}
 
\def\boxeqn#1{\vcenter{\vbox{\hrule\hbox{\vrule\kern3pt\vbox{\kern3pt
	\hbox{${\displaystyle #1}$}\kern3pt}\kern3pt\vrule}\hrule}}}

\def\lform{\hbox{$\sqcup$}\llap{\hbox{$\sqcap$}}}

\def\half{{1\over 2}}

\def\bar{\overline}
\def\({\left(}
\def\){\right)}



\def\qed{\hbox{\hskip 3pt
\vbox{\hrule\hbox to 7pt{\vrule height 7pt\hfill\vrule}
\hrule}}\hskip3pt}

\overfullrule=0pt\relax

\frenchspacing

\newread\instream \openin\instream= label.defs
\ifeof\instream \message{No labels in advance yet. Wait till next pass.}
\else \closein\instream \input label.defs
\fi
\writedefs

\def\arXiv:#1].{\hepthStrip#1 \nil}
\def\hepthStrip#1 #2\nil{\href{http://arxiv.org/abs/#1}{arXiv:#1 #2\unskip}].}

\input epsf

\def\centretable#1{ \hbox to \hsize {\hfill\vbox{
                    \offinterlineskip \tabskip=0pt \halign{#1} }\hfill} }

\preprint UUITP-37/21

\title 10D Super-Yang-Mills Scattering Amplitudes From Its Pure Spinor Action

\author Maor Ben-Shahar\email{\dagger}{benshahar.maor@physics.uu.se} and Max Guillen\email{\ddagger}{max.guillen@physics.uu.se}

\address
Department of Physics and Astronomy, 75108 Uppsala, Sweden

\abstract

Using the pure spinor master action for 10D super-Yang-Mills in the gauge $b_{0}V = Q\Xi$, tree-level scattering amplitudes are calculated through the perturbiner method, and shown to match those obtained from pure spinor CFT techniques. We find kinematic numerators made of nested $b$-ghost operators, and show that the Siegel gauge condition $b_{0}V = 0$ gives rise to color-kinematics duality satisfying numerators whose Jacobi identity follows from the Jacobi identity of a kinematic algebra.

\Date {August 2021}


\lref\form{
J.~A.~M.~Vermaseren,
``New features of FORM,''
[arXiv:0010025 [math-ph]]
}

\lref\PSS{
C.~R.~Mafra,
``PSS: A FORM Program to Evaluate Pure Spinor Superspace Expressions,''
[arXiv:1007.4999 [hep-th]]
}

\lref\Siegelnongo{
W.~Siegel and M.~Rocek,
``ON OFF-SHELL SUPERMULTIPLETS,''
Phys. Lett. B {\bf 105}, 275-277 (1981).
}

\lref\taylorsugra{
J.~G.~Taylor,
``A No Go Theorem for Off-shell Extended Supergravities,''
J. Phys. A {\bf 15}, 867 (1982).
}

\lref\Rivellesnongo{
V.~O.~Rivelles and J.~G.~Taylor,
``Off-shell No Go Theorems for Higher Dimensional Supersymmetries and Supergravities,''
Phys. Lett. B {\bf 121}, 37-42 (1983).
}

\lref\Nilssonaux{
B.~E.~W.~Nilsson,
``Pure Spinors as Auxiliary Fields in the Ten-dimensional Supersymmetric Yang-Mills Theory,''
Class. Quant. Grav. {\bf 3}, L41 (1986).
}

\lref\Cederwallthreed{
M.~Cederwall,
``Superfield actions for N=8 and N=6 conformal theories in three dimensions,''
JHEP {\bf 10}, 070 (2008).
[arXiv:0809.0318 [hep-th]].
}

\lref\Cederwallborninfeld{
M.~Cederwall and A.~Karlsson,
``Pure spinor superfields and Born-Infeld theory,''
JHEP {\bf 11}, 134 (2011).
[arXiv:1109.0809 [hep-th]].
}

\lref\Berkovitsn{
N.~Berkovits and N.~Nekrasov,
``Multiloop superstring amplitudes from non-minimal pure spinor formalism,''
JHEP {\bf 12}, 029 (2006).
[arXiv:0609012 [hep-th]].
}

\lref\nathanictp{
N.~Berkovits,
``ICTP lectures on covariant quantization of the superstring,''
ICTP Lect. Notes Ser. {\bf 13}, 57-107 (2003).
[arXiv:hep-th/0209059 [hep-th]].
}

\lref\Cederwallsugra{
M.~Cederwall,
``D=11 supergravity with manifest supersymmetry,''
Mod. Phys. Lett. A {\bf 25}, 3201-3212 (2010).
[arXiv:1001.0112 [hep-th]].
}

\lref\Cederwalloverview{
M.~Cederwall,
``Pure spinor superfields -- an overview,''
Springer Proc. Phys. {\bf 153}, 61-93 (2014).
[arXiv:1307.1762 [hep-th]].
}

\lref\Berkovitsstring{
N.~Berkovits,
``Super Poincare covariant quantization of the superstring,''
JHEP {\bf 04}, 018 (2000).
[arXiv:0001035 [hep-th]].
}

\lref\Berkovitsvanishing{
N.~Berkovits,
``Multiloop amplitudes and vanishing theorems using the pure spinor formalism for the superstring,''
JHEP {\bf 09}, 047 (2004).
[arXiv:hep-th/0406055 [hep-th]].
}

\lref\Siegelgauge{
W.~Siegel,
``Covariantly Second Quantized String. 2.,''
Phys. Lett. B {\bf 149}, 157 (1984).
}

\lref\Siegelintroduction{
W.~Siegel,
``Introduction to string field theory,''
Adv. Ser. Math. Phys. {\bf 8}, 1-244 (1988).
[arXiv:0107094 [hep-th]].
}

\lref\Cederwallsupergeometry{
M.~Cederwall,
``From supergeometry to pure spinors,''
[arXiv:1012.3334 [hep-th]].
}

\lref\Karlsson{
A.~Karlsson,
``Ultraviolet divergences in maximal supergravity from a pure spinor point of view,''
JHEP {\bf 04}, 165 (2015).
[arXiv:1412.5983 [hep-th]].
}

\lref\Bjornssonone{
J.~Bjornsson and M.~B.~Green,
``5 loops in 24/5 dimensions,''
JHEP {\bf 08}, 132 (2010).
[arXiv:1004.2692 [hep-th]].
}

\lref\Bjornssontwo{
J.~Bjornsson,
``Multi-loop amplitudes in maximally supersymmetric pure spinor field theory,''
JHEP {\bf 01}, 002 (2011).
[arXiv:1009.5906 [hep-th]].
}

\lref\Mafraoneloopone{
C.~R.~Mafra and O.~Schlotterer,
``Towards the n-point one-loop superstring amplitude. Part I. Pure spinors and superfield kinematics,''
JHEP {\bf 08}, 090 (2019).
[arXiv:1812.10969 [hep-th]].
}

\lref\Mafraonelooptwo{
C.~R.~Mafra and O.~Schlotterer,
``Towards the n-point one-loop superstring amplitude. Part II. Worldsheet functions and their duality to kinematics,''
JHEP {\bf 08}, 091 (2019).
[arXiv:1812.10970 [hep-th]].
}

\lref\Mafratwoloop{
E.~D'Hoker, C.~R.~Mafra, B.~Pioline and O.~Schlotterer,
``Two-loop superstring five-point amplitudes. Part I. Construction via chiral splitting and pure spinors,''
JHEP {\bf 08}, 135 (2020).
[arXiv:2006.05270 [hep-th]].
}

\lref\Mafrathreeloop{
H.~Gomez and C.~R.~Mafra,
``The closed-string 3-loop amplitude and S-duality,''
JHEP {\bf 10}, 217 (2013).
[arXiv:1308.6567 [hep-th]].
}

\lref\harnadshnider{
J.~P.~Harnad and S.~Shnider,
``CONSTRAINTS AND FIELD EQUATIONS FOR TEN-DIMENSIONAL SUPERYANG-MILLS THEORY,''
Commun. Math. Phys. {\bf 106}, 183 (1986).
}

\lref\Annaone{
M.~Cederwall and A.~Karlsson,
``Loop amplitudes in maximal supergravity with manifest supersymmetry,''
JHEP {\bf 03}, 114 (2013).
[arXiv:1212.5175 [hep-th]].
}

\lref\Annatwo{
A.~Karlsson,
``Ultraviolet divergences in maximal supergravity from a pure spinor point of view,''
JHEP {\bf 04}, 165 (2015).
[arXiv:1412.5983 [hep-th]].
}

\lref\Cederwallfourd{
M.~Cederwall,
``An off-shell superspace reformulation of D=4, N=4 super-Yang-Mills theory,''
Fortsch. Phys. {\bf 66}, no.1, 1700082 (2018).
[arXiv:1707.00554 [hep-th]].
}

\lref\Cederwallsixd{
M.~Cederwall,
``Pure spinor superspace action for D = 6, N = 1 super-Yang-Mills theory,''
JHEP {\bf 05}, 115 (2018).
[arXiv:1712.02284 [hep-th]].
}

\lref\Mafraone{
C.~R.~Mafra, O.~Schlotterer and S.~Stieberger,
``Complete N-Point Superstring Disk Amplitude I. Pure Spinor Computation,''
Nucl. Phys. B {\bf{873}}, 419-460 (2013).
[arXiv:1106.2645 [hep-th]].
}

\lref\Mafratwo{
C.~R.~Mafra and O.~Schlotterer,
``Multiparticle SYM equations of motion and pure spinor BRST blocks,''
JHEP {\bf 07}, 153 (2014).
[arXiv:1404.4986 [hep-th]].
}

\lref\bcjone{
    Z.~Bern, J.~J.~Carrasco, and H.~Johansson,
    ``New Relations for Gauge-Theory Amplitudes,''
    Phys. Rev. D {\bf{78}},  085011 (2008).
   [arXiv:0805.3993 [hep-ph]].
}

\lref\bcjtwo{
    Z.~Bern, J.~J.~Carrasco, and H.~Johansson,
    ``Perturbative Quantum Gravity as a Double Copy of Gauge Theory,''
    Phys. Rev. Lett. {\bf{105}},  061602 (2010).
   [arXiv:1004.0476 [hep-th]].
}

\lref\bcjreview{
    Z.~Bern, J.~J.~Carrasco, and H.~Johansson,
    ``The Duality Between Color and Kinematics and its Applications,''
    [arXiv:1909.01358 [hep-th]].
}

\lref\nathandynamical{
N.~Berkovits,
``Dynamical twisting and the b ghost in the pure spinor formalism,''
JHEP {\bf 06}, 091 (2013).
[arXiv:1305.0693 [hep-th]].
}

\lref\Berkovitssuperparticle{
N.~Berkovits,
``Covariant quantization of the superparticle using pure spinors,''
JHEP {\bf 09}, 016 (2001).
[arXiv:hep-th/0105050 [hep-th]].
}

\lref\Berkovitsmax{
N.~Berkovits and M.~Guillen,
``Equations of motion from Cederwall's pure spinor superspace actions,''
JHEP {\bf 08}, 033 (2018).
[arXiv:1804.06979 [hep-th]].
}

\lref\Berkovitstop{
N.~Berkovits,
``Pure spinor formalism as an N=2 topological string,''
JHEP {\bf 10}, 089 (2005).
[arXiv:hep-th/0509120 [hep-th]].
}

\lref\max{
M.~Guillen,
``Notes on the 11D pure spinor wordline vertex operators,''
JHEP {\bf 08}, 122 (2020).
[arXiv:2006.06022 [hep-th]].
}

\lref\flavourkinematics{C.~Cheung and C.~Shen,
``Symmetry for Flavor-Kinematics Duality from an Action,''
Phys. Rev. Lett. {\bf{118}}, 121601 (2017).
[arXiv:1612.00868 [hep-th]].
}

\lref\borstencomplicated{
L.~Borsten, H.~Kim, B.~Jur\v{c}o, T.~Macrelli, C.~Saemann and M.~Wolf,
``Double Copy from Homotopy Algebras,''
[arXiv:2102.11390 [hep-th]].
}

\lref\borstenloops{L.~Borsten, B.~Jurco, H.~Kim, T.~Macrelli, C.~Saemann and M.~Wolf,
``Tree-Level Color-Kinematics Duality Implies Loop-Level Color-Kinematics Duality,''
[arXiv:2108.03030 [hep-th]].
}

\lref\cheungnew{
"C.~Cheung and J.~Mangan,
``Covariant Color-Kinematics Duality,''
[arXiv:2108.02276 [hep-th]].
}

\lref\firstnmhvalgebra{
G.~Chen, H.~Johansson, F.~Teng and T.~Wang,
``On the kinematic algebra for BCJ numerators beyond the MHV sector,''
JHEP {\bf 11} 055 (2019).
[arXiv:1906.10683 [hep-th]].
}

\lref\nmhvalgebra{
G.~Chen, H.~Johansson, F.~Teng and T.~Wang,
``Next-to-MHV Yang-Mills kinematic algebra,''
[arXiv:2104.12726 [hep-th]].
}

\lref\tolotti{
M.~Tolotti and S.~Weinzierl,
``Construction of an effective Yang-Mills Lagrangian with manifest BCJ duality,''
JHEP {\bf{07}}, 111 (2013).
[arXiv:1306.2975 [hep-th]].
}

\lref\originalcontactterms{
Z.~Bern, T.~Dennen, Y.~Huang and M.~Kiermaier,
``Gravity as the Square of Gauge Theory,''
Phys. Rev. D {\bf{82}}, 065003 (2010).
[arXiv:1004.0693 [hep-th]].
}

\lref\reiterer{
M.~Reiterer,
``A homotopy BV algebra for Yang-Mills and color-kinematics'',
[arXiv:1912.03110 [math-ph]].
}


\lref\selfdualalgebra{
R.~Monteiro and D.~O'Connell,
``The Kinematic Algebra From the Self-Dual Sector,''
JHEP {\bf{07}}, 007 (2011).
[arXiv:1105.2565 [hep-th]].
}

\lref\bridges{
E.~Bridges and C.~Mafra,
``Algorithmic construction of SYM multiparticle superfields in the BCJ gauge,''
JHEP {\bf 10}, 022 (2019).
[arXiv:1906.12252 [hep-th]].
}

\lref\marco{
M.~Chiodaroli, Q.~Jin and R.~Roiban,
``Color/kinematics duality for general abelian orbifolds of N=4 super Yang-Mills theory,''
JHEP {\bf 01}, 152 (2014).
[arXiv:1311.3600 [hep-th]].
}

\lref\Mafrathree{
S.~Lee, C.~Mafra and O.~Schlotterer,
``Non-linear gauge transformations in $D=10$ SYM theory and the BCJ duality,''
JHEP {\bf 03}, 090 (2016).
[arXiv:1510.08843 [hep-th]].
}

\lref\selivanov{
K.~.G.~Selivanov,
``On tree form-factors in (supersymmetric) Yang-Mills theory,''
Commun. Math. Phys. {\bf 208}, 671 (2000).
[arXiv:9809046 [hep-th]].
}

\lref\berendsgiele{
F.~A.~Berends and W.~T.~Giele,
``Multiple Soft Gluon Radiation in Parton Processes,''
Nucl. Phys. B. {\bf 313}, 595 (1989).
}

\lref\Henrikmaor{
M.~Ben-Shahar and H.~Johansson, {\it to appear.}
}

\lref\cheung{
C.~Cheung and J.~Mangan,
``Covariant Color-Kinematics Duality,''
[arXiv:2108.02276 [hep-th]].
}


\lref\bernone{
Z.~Bern, J.~J.~Carrasco, L.~Dixon, H.~Johansson and R.~Roiban,
``Simplifying Multiloop Integrands and Ultraviolet Divergences of Gauge Theory and Gravity Amplitudes,''
Phys. Rev. D. {\bf 85}, 105014 (2012).
[arXiv:1201.5366 [hep-th]].
}

\lref\berntwo{
Z.~Bern, J.~J.~Carrasco, W.~M.~Chen, A.~Edison, H.~Johansson, J.~Parra-Martinez, R.~Roiban and M.~Zeng,
``Ultraviolet Properties of ${\cal{N}} = 8$ Supergravity at Five Loops,''
Phys. Rev. D {\bf 98}, 8, 086021 (2018).
[arXiv:1804.09311 [hep-th]].
}


\lref\tsimpisone{
    G.~Policastro and D.~Tsimpis,
    ``R**4, purified,''
    Class. Quant. Grav. {\bf 23} 4753 (2006).
    [arXiv:0603165 [hep-th]].
}

\lref\chang{
    C-M.~Chang, Y-H.~Lin, Y.~Wang and X.~Yin,
     ``Deformations with Maximal Supersymmetries Part 2: Off-shell Formulation,''
     JHEP {\bf 04} 171 (2016).
     [arXiv:1403.0709 [hep-th]].
}

\lref\Mafrarefone{
    C.~R.~Mafra,
    ``Towards Field Theory Amplitudes From the Cohomology of Pure Spinor Superspace,''
    JHEP {\bf 11} 096 (2010).
    [arXiv:1007.3639 [hep-th]].
}
\lref\Mafrareftwo{
    C.~R.~Mafra, O.~Schlotterer, S.~Stieberger, and D.~Tsimpis,
    ``A recursive method for SYM n-point tree amplitudes,''
    Phys. Rev. D {\bf 83} 126012 (2011).
    [arXiv:1012.3981 [hep-th]].
}
\lref\Mafrarefthree{
    C.~R.~Mafra, O.~Schlotterer and S.~Stieberger,
    ``Explicit BCJ Numerators from Pure Spinors,''
    JHEP {\bf 07} 092 (2011).
    [arXiv:1104.5224 [hep-th]].
}
\lref\Mafrareffour{
    C.~R.~Mafra and O.~Schlotterer,
    ``Cohomology foundations of one-loop amplitudes in pure spinor superspace,''
    [arXiv:1408.3605 [hep-th]].
}
\lref\Mafrareffive{
    C.~R.~Mafra and O.~Schlotterer,
    ``Solution to the nonlinear field equations of ten dimensional supersymmetric Yang-Mills theory,''
    Phys. Rev. D {\bf 92} 066001 (2015).
    [arXiv:1501.05562 [hep-th]].
}
\lref\Mafrarefsixn{
    C.~R.~Mafra and O.~Schlotterer,
    ``Berends-Giele recursions and the BCJ duality in superspace and components,''
    JHEP {\bf 03} 097 (2016).
    [arXiv:1510.08846 [hep-th]].
}

\lref\copenhagengroup{
     N.~E.~J.~Bjerrum-Bohr, P.~H.~Poul, T.~Sondergaard and P.~Vanhove,
     ``The Momentum Kernel of Gauge and Gravity Theories,''
     JHEP {\bf 01} 001 (2011).
    [arXiv:1010.3933 [hep-th]].
}

\lref\alexandfei{
     A.~Edison and F.~Teng,
     ``Efficient Calculation of Crossing Symmetric BCJ Tree Numerators,''
     JHEP {\bf 12} 138 (2020).
    [arXiv:2005.03638 [hep-th]].
}

\font\mbb=msbm10 
\newfam\bbb
\textfont\bbb=\mbb

\def\startcenter{%
  \par
  \begingroup
  \leftskip=0pt plus 1fil
  \rightskip=\leftskip
  \parindent=0pt
  \parfillskip=0pt
}
\def\stopcenter{%
  \par
  \endgroup
}

\listtoc
\writetoc
\filbreak

\newsec Introduction

\seclab\secone

\noindent
It has been known for a long time that maximally supersymmetric theories lack a manifestly Lorentz covariant Lagrangian formulation in ordinary superspace \refs{\Siegelnongo,\taylorsugra,\Rivellesnongo},
making it more difficult to exploit supersymmetry in amplitudes computations. However, things change when superspace is extended \Nilssonaux.
Indeed, non-minimal pure spinor variables have been used in \refs{\Cederwallthreed,\Cederwallsugra,\Cederwallborninfeld,\Cederwalloverview} for constructing manifestly supersymmetric actions for several maximally supersymmetric theories like 10D super-Yang-Mills, 10D super-Born-Infeld and 11D supergravity. This makes pure spinor quantum field theory a promising approach for evaluating scattering amplitudes in an elegant way. This should not be a surprise at all. In fact, the pure spinor formalism for superstrings \Berkovitsstring\ is nowadays arguably the most powerful and efficient framework for computing string scattering amplitudes as compared to the traditional Ramond-Neveu-Schwarz and Green-Schwarz formalisms \refs{\Mafraone,\Mafraoneloopone,\Mafraonelooptwo,\Mafratwoloop,\Mafrathreeloop}.

\medskip
Such pure spinor field theory actions have been constructed from a single pure spinor superfield, which exhibits the field-antifield symmetry of the corresponding Batalin-Vilkovisky descriptions of the theories in study. Since standard gauge-fixing fermions are incompatible with this symmetry, alternative gauge-fixing conditions are necessary for the computation of scattering amplitudes. Using inspiration from string field theory, the Siegel gauge condition $b_{0}V = 0$ \refs{\Siegelgauge, \Siegelintroduction}, has been proposed in \refs{\Cederwalloverview, \Cederwallsupergeometry} as a natural gauge choice. Although no explicit computations have been done so far in this pure field theory setting, power counting arguments have been used to discuss the ultraviolet behaviour of 10D super-Yang-Mills and 11D supergravity \refs{\Bjornssonone,\Bjornssontwo,\Annaone,\Annatwo}.

\medskip
On the other hand, recent progress in the study of scattering amplitudes has revealed new structures previously hidden from Lagrangian formulations of field theories. The BCJ duality between color-and kinematics \refs{\bcjone,\bcjtwo,\bcjreview} states that given an amplitude formulated as a sum over cubic diagrams, 
\eqnn \auxone
$$ \eqalignno{
A &= \sum_{i\in\Gamma_i}{c_i n_i \over D_i} \ ,
 & \auxone
}
$$
it is possible to find \refs{\copenhagengroup,\Mafrathree} representations of the kinematic numerators such that they obey the same relations as the color factors
\eqnn \auxtwo
$$ \eqalignno{
c_i + c_j +c_k &= 0 \Rightarrow n_i + n_j + n_k = 0 \ . 
& \auxtwo
}
$$
Having obtained such numerators, it is possible to ``double-copy'' them by replacing $c_i\to n_i$, obtaining amplitudes from a theory of gravity, dramatically simplifying the computation of gravity observables \refs{\bernone,\berntwo}. Since the identities obeyed by the color factors are inherited from the Jacobi identity of the color algebra, it is tempting to ask if a similar algebra exists for kinematics. Indeed, algebraic origins of this duality have been observed in certain sectors of Yang-Mills theory \refs{\selfdualalgebra,\flavourkinematics,\firstnmhvalgebra,\nmhvalgebra}, but the general case is still not fully understood, appearing to require an infinite number of auxiliary fields \refs{\originalcontactterms,\tolotti, \borstencomplicated,\borstenloops,\reiterer} whose job is to modify the numerators by generalized gauge transformations. 

\medskip
In this paper we study the 10D super-Yang-Mills pure spinor action subject to two different gauge-fixing conditions. The first gauge choice is a relaxation of the standard Siegel gauge condition, namely, $b_{0}V = Q\Xi$, where $b_{0}$ is the quantum-mechanical operator version of the pure spinor $b$-ghost, and $\Xi$ is some superfield. The usual unintegrated vertex operators $V = \lambda^{\alpha}A_{\alpha}$, widely used in superstring scattering amplitudes, are explicitly shown to satisfy such a condition. We then compute the corresponding equations of motion in pure spinor superspace, and apply the perturbiner method to compute tree-level scattering amplitudes from simple contractions of Berends-Giele currents. Our numerators take the form of nested $b$-ghosts acting on external states, for example the half-ladder numerator at five points is $\langle b_0(b_0(V_1 V_2)V_3)V_4V_5\rangle$. We explicitly calculate such expressions for 4-point and 5-point amplitudes by making use of the so-called physical operators introduced in \Cederwallborninfeld, in the process collecting useful identities for these operators. The amplitudes thus obtained are shown to coincide with those found from open pure spinor superstrings in the particle-limit \refs{\Mafraone,\Mafrareftwo ,\Mafrareffive}. 
We also study the Siegel gauge, $b_0 V = 0$, in which the vertex operators are dependant on non-minimal variables. In this scenario, nilpotency of the $b$-ghost is shown to imply color-kinematics duality off-shell.
The nested $b$-ghosts define a Poisson bracket, whose Jacobi identity is the kinematic Jacobi identity. The dual Lie algebra to this Poisson algebra is an algebra of diffeomorphisms that preserve the Siegel gauge. Such a gauge appears to be the only one that allows for both manifest crossing symmetry and color-kinematics duality.

\medskip
The paper is organized as follows. In section 2, we discuss the 10D pure spinor superparticle and its BRST-cohomology. Non-minimal pure spinors are then introduced in order to construct both well-defined action principles, and the $b$-ghost which satisfies $\{Q, b\} = {P^{2}\over 2}$. In section 3, we discuss two approaches for computing tree amplitudes, namely, Feynman diagrams and Berends-Giele recursion relations. In section 4, we apply the systematics of the latter and compute N-point scattering amplitudes in 10D super-Yang-Mills when external states satisfy $b_{0}V = Q\Xi$. The stronger condition $b_{0}V = 0$ is then discussed and shown to manifestly reproduce color-kinematics duality. Calculations of different expressions involving nested $b$-ghosts are explicitly carried out in Appendix A. Finally, we close with discussions and further research directions in section 5.

\medskip

\newsec Pure Spinor Action Of 10D Super-Yang-Mills 

\seclab\sectwo

\noindent
In this section we review the minimal 10D pure spinor superparticle. After introducing non-minimal variables, a well-defined pure spinor measure is constructed as well as a composite operator $b$, the so-called $b$-ghost, satisfying $\{Q, b\} = {P^{2}\over 2}$.

\subsec 10D Pure Spinor Superparticle

\subseclab\sectwoone

\noindent
The 10D pure spinor superparticle \Berkovitssuperparticle\ is defined by the action
\eqnn \tendparticle
$$ \eqalignno{
S &= \int d\tau \,\bigg[P_{m}\partial_{\tau}X^{m} + p_{\alpha}\partial_{\tau}\theta^{\alpha} + w_{\alpha}\partial_{\tau}\lambda^{\alpha} - \half P^{2}\bigg] \ , & \tendparticle
}
$$
and the BRST operator
\eqnn \brsttendparticle
$$ \eqalignno{
Q &= \lambda^{\alpha}d_{\alpha} \ , & \brsttendparticle
}
$$
where we are using letters from the beginning/middle of the Greek/Latin alphabet to denote $SO(1,9)$ spinor/vector indices. Furthermore, $(X^{m}, \theta^{\alpha})$ stand for the superspace coordinates, and $(P_{m}, p_{\alpha})$ denote their respective conjugate momenta. The variable $\lambda^{\alpha}$ is a pure spinor satisfying $\lambda\gamma^{m}\lambda = 0$, and $w_{\alpha}$ is its respective conjugate momentum, which is defined up to the gauge transformation $\delta w_{\alpha} = (\gamma^{m}\lambda)_{\alpha}\rho_{m}$, for any $\rho_{m}$. The Green-Schwarz fermionic constraint $d_{\alpha}$ in \brsttendparticle\ is defined as usual: $d_{\alpha} = p_{\alpha} + \half (\gamma^{m}\theta)_{\alpha}P_{m}$, and it commutes with the supersymmetry generators: $q_{\alpha} = p_{\alpha} - \half (\gamma^{m}\theta)_{\alpha}P_{m}$. Finally, $(\gamma^{m})_{\alpha\beta}$, $(\gamma^{m})^{\alpha\beta}$ denote the familiar $SO(1,9)$ Pauli matrices satisfying $(\gamma^{(m})_{\alpha\beta}(\gamma^{n)})^{\beta\delta} = \eta^{mn}\delta^{\delta}_{\alpha}$.

\medskip
The action \tendparticle\ is invariant under Super-Poincare transformations as well as under the global symmetry generated by $J = -\lambda^{\alpha}w_{\alpha}$. The charge associated to the current $J$ will be referred to as ghost number, and thus $\lambda^{\alpha}$, $w_{\alpha}$ carry ghost numbers $1$, $-1$, respectively. As is well known, the Hilbert space of the superparticle \tendparticle\ will be described by the cohomology of the BRST operator \brsttendparticle, which can be conveniently separated into different ghost number sectors. Indeed, it has been computed from different methods that there only exists non-trivial cohomology up to ghost number 3. Schematically,
\eqnn \cohomology
$$
\eqalignno{
\Psi(x,\theta,\lambda) &= \Psi^{(0)}(x,\theta,\lambda) + \Psi^{(1)}(x,\theta,\lambda) + \Psi^{(2)}(x,\theta,\lambda) + \Psi^{(3)}(x,\theta,\lambda) \ , & \cohomology
}
$$
where the superscript in $\Psi^{(i)}$ stands for the ghost number sector which $\Psi^{(i)}$ belongs to. In this manner, the pure spinor superfields $\Psi^{(0)}$, $\Psi^{(1)}$, $\Psi^{(2)}$, $\Psi^{(3)}$ were found to describe the gauge symmetry ghosts, physical fields, antifields, and ghost antifields of 10D super-Yang-Mills, respectively. Let us illustrate this with the ghost number one sector $\Psi^{(1)}(x,\theta,\lambda) = \lambda^{\alpha}A_{\alpha}(x,\theta)$. This superfield is subject to the physical state conditions
\eqnn \physicalconditions
$$ \eqalignno{
Q\Psi^{(1)} &= 0 \ , \ \ \ \ \ \delta \Psi^{(1)} = Q\Lambda  \ , & \physicalconditions
}
$$
where $\Lambda$ is an arbitrary gauge superfield. The implications of \physicalconditions\ on $A_{\alpha}(x,\theta)$ read
\eqnn \physicalconditionsona
$$ \eqalignno{
(\gamma^{mnpqr})^{\alpha\beta}D_{\alpha}A_{\beta} &= 0 \ , \ \ \ \ \ \delta A_{\alpha} = D_{\alpha}\Lambda \ , & \physicalconditionsona
}
$$
where $D_{\alpha} = \partial_{\alpha} + \half (\gamma^{m}\theta)_{\alpha}\partial_{m}$ is the usual supersymmetric derivative. Eqns. \physicalconditionsona\ are nothing but the superspace equations of motion of linearized 10D super-Yang-Mills. They can easily be solved in the so-called Harnad-Shnider gauge \harnadshnider, which imposes $\theta^{\alpha}A_{\alpha} = 0$. In this gauge, $A_{\alpha}(x,\theta)$ takes the form \tsimpisone
\eqnn \aalpha
$$ \eqalignno{
A_{\alpha}(x,\theta) &= \half (\gamma^{m}\theta)_{\alpha}a_{m}(x) - {1\over 3}(\gamma^{m}\theta)_{\alpha}(\theta\gamma_{m}\chi(x)) - {1 \over 16}(\gamma_{p}\theta)_{\alpha}(\theta\gamma^{mnp}\theta)\partial_{m}a_{n}(x)\cr 
& + {1 \over 60}(\gamma_{p}\theta)_{\alpha}(\theta\gamma^{mnp}\theta)(\theta\gamma_{m}\partial_{n}\chi(x))
 + \ldots & \aalpha
}
$$
where $\ldots$ stands for higher derivative terms of $a_{m}$, $\chi^{\alpha}$. The field $a_{m}$ in \aalpha\ satisfies the relations $\partial^{m}a_{m} = 0$, $\delta a_{m} = \partial_{m}s$ for any $s$, and $\chi^{\alpha}$ satisfies the equation $(\gamma^{m})_{\alpha\beta}\partial_{m}\chi^{\beta} = 0$, so that $a_{m}$, $\chi^{\alpha}$ are identified with the 10D super-Yang-Mills gluon and gluino fields, respectively. Similar arguments apply to the other ghost sectors, and so \cohomology\ indeed describes the Batalin-Vilkovisky formulation of linearized 10D super-Yang-Mills.

\medskip
The top scalar cohomology of $Q$ can be used to define manifestly supersymmetric correlators \Berkovitsstring. Such a measure has played a fundamental role in the computation of superstring scattering amplitudes. Its explicit form reads
\eqnn \minimalmeasure
$$ \eqalignno{
\langle (\lambda\gamma^{m}\theta)(\lambda\gamma^{n}\theta)(\lambda\gamma^{p}\theta)(\theta\gamma_{mnp}\theta)\rangle &= 1 \ . &\minimalmeasure
}
$$
One might then naively use this measure for construcing a manifestly supersymmetric action reproducing eqns. \physicalconditions. However, the measure \minimalmeasure\ is degenerate \refs{\Cederwalloverview,\nathanictp}\ and so it is not adequate for such a purpose. This problem is solved by introducing non-minimal pure spinor variables. This is what we do next.

\subsec Non-Minimal Variables

\subseclab\sectwotwo

The pair of conjugate variables $(\bar{\lambda}_{\alpha}, \bar{w}^{\alpha})$, $(r_{\alpha}, s^{\alpha})$, where $\bar{\lambda}_{\alpha}$ is a pure spinor satisfying $\bar{\lambda}\gamma^{m}\bar{\lambda} = 0$ and $r_{\alpha}$ is a fermionic variable satisfying $\bar{\lambda}\gamma^{m}r = 0$, can be introduced into the model \tendparticle\ through the quartet argument, that is
\eqnn \nonminimalaction
$$ \eqalignno{
S &= \int d\tau \,\bigg[P_{m}\partial_{\tau}X^{m} + p_{\alpha}\partial_{\tau}\theta^{\alpha} + w_{\alpha}\partial_{\tau}\lambda^{\alpha} + \bar{w}^{\alpha}\partial_{\tau}\bar{\lambda}_{\alpha} + s^{\alpha}\partial_{\tau}r_{\alpha} - \half P^{2}\bigg] \ ,& \nonminimalaction
}
$$
and,
\eqnn \qnonminimal
$$ \eqalignno{
Q &= Q_{0} + r_{\alpha}\bar{w}^{\alpha} \ , & \qnonminimal
}
$$
so that the BRST-cohomology does not change \Berkovitstop. Indeed, it is not hard to see it is always possible to find a representative element in the cohomology of \qnonminimal\ which is independent of the non-minimal variables. These variables allow us to write down a well-defined measure for formulating action principles in pure spinor superspace. Such a measure takes the form $[dZ] = [d\lambda][d\bar{\lambda}][dr]$, where
\eqnn \nonminimalmeasureone
\eqnn \nonminimalmeasuretwo
\eqnn \nonminimalmeasurethree
$$ \eqalignno{
[d\lambda]\lambda^{\beta}\lambda^{\delta}\lambda^{\gamma} &= (\epsilon T^{-1})^{\beta\delta\gamma}_{\alpha_{1}\ldots \alpha_{11}}d\lambda^{\alpha_{1}}\ldots d\lambda^{\alpha_{11}} & \nonminimalmeasureone\cr
[d\bar{\lambda}]\bar{\lambda}_{\beta}\bar{\lambda}_{\delta}\bar{\lambda}_{\gamma} &= (\epsilon T)_{\beta\delta\gamma}^{\alpha_{1}\ldots \alpha_{11}}d\bar{\lambda}_{\alpha_{1}}\ldots d\bar{\lambda}_{\alpha_{11}} & \nonminimalmeasuretwo\cr
[dr] &= (\epsilon T^{-1})^{\beta\delta\gamma}\bar{\lambda}_{\beta}\bar{\lambda}_{\delta}\bar{\lambda}_{\gamma}\big({\partial \over \partial r_{\alpha_{1}}}\big)\ldots \big({\partial \over \partial r_{\alpha_{11}}}\big) & \nonminimalmeasurethree
}
$$
and the Lorentz invariant tensors $(\epsilon T^{-1})^{\beta\gamma\delta}_{\alpha_{1}\ldots \alpha_{11}}$, $(\epsilon T)_{\beta\gamma\delta}^{\alpha_{1}\ldots \alpha_{11}}$ are symmetric and gamma-traceless in $(\beta,\delta,\gamma)$ and antisymmetric in $[\alpha_{1}, \ldots, \alpha_{11}]$. Their explicit forms can be found in \refs{\Berkovitsstring,\Berkovitstop,\Berkovitsvanishing}. For instance, up to an overall normalization factor
\eqnn \epsilont
$$ \eqalignno{
(\epsilon T)^{\alpha_{1}\ldots \alpha_{11}}_{\beta\gamma\delta} &= \epsilon^{\alpha_{1}\ldots \alpha_{16}}(\gamma^{m})_{\alpha_{12}\eta}(\gamma^{n})_{\alpha_{13}\epsilon}(\gamma^{p})_{\alpha_{14}\kappa}(\gamma_{mnp})_{\alpha_{15}\alpha_{16}}\bigg[\delta^{\eta}_{(\beta}\delta^{\epsilon}_{\gamma}\delta^{\kappa}_{\delta)} - {1 \over 40}(\gamma^{q})_{(\beta\gamma}\delta^{\eta}_{\delta)}(\gamma_{q})^{\epsilon\kappa}\bigg]\cr
& & \epsilont
}
$$

\medskip
The linearized 10D super-Yang-Mills action then reads 
\eqnn \nonminimalsym
$$ \eqalignno{
S &= \int d^{10}x d^{16}\theta [dZ]{\cal{N}}\bigg(\half\Psi Q\Psi\bigg) \ , &\nonminimalsym
}
$$
where ${\cal{N}} = e^{-\{Q, \xi\}}$ with $\xi = \bar{\lambda}\theta$, is a regularization factor which prevents the appearance of undetermined expressions as a result of the zero mode integrations of non-compact bosonic and fermionic variables, and $\Psi$ is a pure spinor superfield which generically depends on non-minimal variables. Due to the existence of the operator $\tilde{\xi} = {\bar{\lambda}\theta\over \lambda\bar{\lambda} + r\theta}$ satisfying $Q\tilde{\xi} = 1$, pure spinor integrands as that in \nonminimalsym\ will be restricted to diverge slower than $\lambda^{-8}\bar{\lambda}^{-11}$ \Berkovitstop. Integrands diverging faster than $\lambda^{-8}\bar{\lambda}^{-11}$ require a different regularization scheme \Berkovitsn, which will not be discussed in this paper. It is straightforward to show that the pure spinor action \nonminimalsym\ indeed reproduces eqns. \physicalconditions. In order to introduce interactions, the field-antifield structure, inherently described by the pure spinor superfield $\Psi$, can be exploited by introducing the so-called pure spinor antibracket, defined via
\eqnn \antibracket
$$ \eqalignno{
(A, B) &= \int  {\delta_{R} A  \over \delta \Psi(Z)}[dZ] {\delta_{L} B \over \delta \Psi(Z)} \ ,
& \antibracket
}
$$
where $L$, $R$ denote left, right derivatives, respectively. Then, a pure spinor master action must satisfy
\eqnn \masteraction
$$ \eqalignno{
(S, S) &= 0 \ . &\masteraction
}
$$
As can easily be demonstrated, the action \nonminimalsym\ satisfies the master equation \masteraction. The simplest deformation of \nonminimalsym\ which does not explicitly involve non-minimal variables and satisfies \masteraction\ reads
\eqnn \nonabelian
$$ \eqalignno{
S &= \int\,d^{10}xd^{16}\theta [dZ]{\cal{N}}\, {\rm Tr}\bigg(\half \Psi Q\Psi + {g \over 3}\Psi\Psi\Psi\bigg) \ ,  &\nonabelian
}
$$
where ${\rm Tr}$ is the trace taken over gauge group generator matrices, and  $g$ is the coupling constant. The equations of motion and gauge transformations following from \nonabelian\ can readily be computed to be
\eqnn \eomgt
$$ \eqalignno{
Q\Psi + g\Psi\Psi &= 0 \ , \ \ \ \ \ \ \delta \Psi = Q\Lambda  + g[\Psi,\Lambda] \ , & \eomgt
}
$$
where $\Lambda$ is an arbitrary gauge superfield. Eqns. \eomgt\ imply that the ghost number one sector $\Psi^{(1)} = \lambda^{\alpha}A_{\alpha}$ satisfies
\eqnn \eomsuperspace
$$ \eqalignno{
(\gamma^{mnpqr})^{\alpha\beta}(D_{\alpha}A_{\beta} + g A_{\alpha}A_{\beta}) &= 0 \ , \ \ \ \ \ \ \delta A_{\alpha} = D_{\alpha}\Lambda +  g[A_{\alpha},\Lambda]\ .
& \eomsuperspace
}
$$
These equations are nothing but the superspace constraints describing 10D non-abelian super-Yang-Mills. Therefore, \eomgt\ can be viewed as the superspace equations of motion describing full 10D super-Yang-Mills in its antifield formulation.

\subsec The $b$-ghost

\subseclab\sectwothree

In the non-minimal pure spinor worldline formalism, there exists a composite operator $b$, also known as the $b$-ghost, which obeys the relation
\eqnn \bghost
$$ \eqalignno{
\{Q, b\} &= {P^{2} \over 2} \ . &\bghost
}
$$
Its explicit form reads
\eqnn \explicitbghost
$$ \eqalignno{
b = {(\bar{\lambda}\gamma^{m}d) \over 2(\lambda\bar{\lambda})}P_{m} +& {(\bar{\lambda}\gamma^{mnp}r)[-(d\gamma_{mnp}d) + 24N_{mn}P_{p}]\over 192 (\lambda\bar{\lambda})^{2}} - {(r\gamma^{mnp}r)(\bar{\lambda}\gamma_{m}d)N_{np} \over 16(\lambda\bar{\lambda})^{3}} \cr
& - {(r\gamma^{mnp}r)(\bar{\lambda}\gamma^{pqr}r)N_{mn}N_{qr}\over 128(\lambda\bar{\lambda})^{4}} \ , & \explicitbghost
}
$$
where $N^{mn} = \half (\lambda\gamma^{mn}w)$. This operator has been notably simplified in \nathandynamical\ to a quadratic polynomial in a fermionic vector $\bar{\Gamma}^{m}$, which is linear in $d_{\alpha}$ and $N^{mn}$. Although this simplified form is useful to check several properties satisfied by the $b$-ghost, we will use an alternative expression which makes use of the so-called physical operators, introduced in \Cederwallborninfeld\ in the search for deformations of the 10D super-Yang-Mills action. These operators are defined by the relations
\eqnn \physicaloperatorsone
\eqnn \physicaloperatorstwo
\eqnn \physicaloperatorsthree
\eqnn \physicaloperatorsfour
$$ \eqalignno{
[Q, {\bf{A}}_{\alpha}] &= -d_{\alpha} + (\gamma^{m}\lambda)_{\alpha}{\bf{A}}_{m}\ , & \physicaloperatorsone\cr
\{Q, {\bf{A}}_{m}\} &= P_{m} + (\lambda\gamma_{m}{\bf{W}}) \ ,& \physicaloperatorstwo \cr
[Q, {\bf{W}}^{\alpha}] &= {1 \over 4}(\gamma^{mn})_{\alpha}{}^{\beta}\lambda^{\alpha}{\bf{F}}_{mn} \ ,& \physicaloperatorsthree\cr
\{Q, {\bf{F}}_{mn}\} &= -2(\lambda\gamma_{[m}\partial_{n]}{\bf{W}})\ , & \physicaloperatorsfour
}
$$
which are reminiscent of the standard 10D super-Yang-Mills equations of motion
\eqnn \tendsym
$$ \eqalignno{
D_{\alpha}A_{\beta} + D_{\beta}A_{\alpha} = (\gamma^{m})_{\alpha\beta}A_{m} \ &, \ \ \ \ D_{\alpha}A_{m} = \partial_{m}A_{\alpha} + (\gamma_{m}W)_{\alpha}\ , \cr
D_{\alpha}W^{\beta} = {1\over 4}(\gamma^{mn})_{\alpha}{}^{\beta}F_{mn} \ &, \ \ \ \ D_{\alpha}F_{mn} = -2(\gamma_{[m}\partial_{n]}W)_{\alpha} \ .& \tendsym
}
$$
Since the physical operators carry negative ghost number, the solution to eqns. \physicaloperatorsone-\physicaloperatorsfour\ will necessarily make use of non-minimal variables. Such a solution is given by
\eqnn \solutionphyisone
\eqnn \solutionphyistwo
\eqnn \solutionphyisthree
\eqnn \solutionphyifour
$$ \eqalignno{
{\bf{A}}_{\alpha} &= {1 \over 4(\lambda\bar{\lambda})}\bigg[ N^{mn}(\gamma_{mn}\bar{\lambda})_{\alpha} + J\bar{\lambda}_{\alpha}\bigg] \ ,& \solutionphyisone \cr
{\bf{A}}_{m} &= {(\bar{\lambda}\gamma_{m}d)\over 2(\lambda\bar{\lambda})} + {(\bar{\lambda}\gamma_{mnp}r)\over 8(\lambda\bar{\lambda})^{2}}N^{np}\ , & \solutionphyistwo \cr
{\bf{W}}^{\alpha} &= {(\gamma^{m}\bar{\lambda})^{\alpha}\over 2(\lambda\bar{\lambda})}\bf{\Delta}_{m}\ ,& \solutionphyisthree\cr
{\bf{F}}_{mn} &= -{(r\gamma_{mn}{\bf{W}})\over 2(\lambda\bar{\lambda})} = {(\bar{\lambda}\gamma_{mn}{}^{p}r)\over 4(\lambda\bar{\lambda})}\bf{\Delta}_{p} \ ,& \solutionphyifour
}
$$
where,
\eqnn \deltadef
$$ \eqalignno{
\bf{\Delta}_{m} &= -P_{m} + {(r\gamma_{m}d)\over 2(\lambda\bar{\lambda})} + {(r\gamma_{mnp}r)\over 8(\lambda\bar{\lambda})^{2}}N^{np} \ . &\deltadef
}
$$
After a few algebraic manipulations, one can show these operators act on the on-shell ghost number one superfield $\Psi^{(1)} = \lambda^{\alpha}A_{\alpha}$ as follows
\eqnn \actionsofphysicalopone
\eqnn \actionsofphysicaloptwo
\eqnn \actionsofphysicalopthree
\eqnn \actionsofphysicalopfour
$$ \eqalignno{
\hat{{\bf{A}}}_{\alpha}\Psi^{(1)} &= A_{\alpha} + (\lambda\gamma_{m})_{\alpha}\sigma_{m} \ ,& \actionsofphysicalopone\cr
\hat{{\bf{A}}}_{m}\Psi^{(1)} &= A_{m} - (\lambda\gamma_{m}\rho) + Q\sigma_{m} \ ,& \actionsofphysicaloptwo\cr
\hat{{\bf{W}}}^{\alpha}\Psi^{(1)} &= W^{\alpha} - Q\rho^{\alpha} + (\gamma^{mn}\lambda)^{\alpha}s_{mn} + \lambda^{\alpha}s \ ,& \actionsofphysicalopthree\cr
\hat{{\bf{F}}}_{mn}\Psi^{(1)} &= F_{mn} - 4Qs_{mn} + (\lambda\gamma_{[m}g_{n]}) + (\lambda\gamma_{mn}g) \ ,& \actionsofphysicalopfour
}
$$
where,
\eqnn \shiftone
$$ \eqalignno{
\sigma_{m} &= -{(\bar{\lambda}\gamma_{m}A)\over 2(\lambda\bar{\lambda})} \ , \ \ \ \ \rho^{\alpha} = {(\gamma^{p}\bar{\lambda})^{\alpha}\over 2(\lambda\bar{\lambda})}(A_{p} + Q\sigma_{p}) \ , \ \ \ \ 
\xi^{\alpha} = W^{\alpha} - Q\rho^{\alpha}\ , \cr
s_{mn} &= {(\bar{\lambda}\gamma_{mn}\xi)\over 8(\lambda\bar{\lambda})} \ , \ \ \ \ 
s = {(\bar{\lambda}\xi) \over 4(\lambda\bar{\lambda})}\ , \ \ \ \ 
r_{mn} = -F_{mn} + 4Qs_{mn} \ ,& \shiftone\cr
 g_{\alpha} &= {(\gamma^{mn}\bar{\lambda})_{\alpha} \over 8(\lambda\bar{\lambda})}r_{mn} - {\bar{\lambda}_{\alpha} \over 2(\lambda\bar{\lambda})} Qs \ , \ \ \ \
 g_{m}^{\alpha} = {(\gamma^{n}\bar{\lambda})^{\alpha} \over (\lambda\bar{\lambda})}r_{nm} \ ,
}
$$
and we are using hatted symbols to denote the operator version of their corresponding unhatted symbols, that is, $(\hat{\bf{A}}_{\alpha}, \hat{\bf{A}}_{m}, \hat{\bf{W}}^{\alpha}, \hat{\bf{F}}_{mn})$ are obtained from 
$({\bf{A}}_{\alpha}, {\bf{A}}_{m}, {\bf{W}}^{\alpha}, {\bf{F}}_{mn})$ 
after performing the replacements: 
\eqnn \replacements
$$
\eqalignno{
& 
P_{m} \rightarrow \partial_{m} \ , \ \ d_{\alpha} \rightarrow D_{\alpha}\ , \ \  w_{\alpha} \rightarrow -\partial_{\lambda^{\alpha}}  \ . &
\replacements
}
$$

\medskip
Using the physical operators \solutionphyisone-\solutionphyifour, the $b$-ghost \explicitbghost\ can then be rewritten in the form
\eqnn \bghostphysicalop
$$ \eqalignno{
b &= \half \bigg[P^{m}{\bf{A}}_{m} - d_{\alpha}{\bf{W}}^{\alpha} - \half N^{mn}{\bf{F}}_{mn}\bigg] \ , & \bghostphysicalop
}
$$
as can easily be demonstrated by directly expanding the right-hand side of \bghostphysicalop. As a check, the use of eqns. \physicaloperatorsone-\physicaloperatorsfour\ allows us to show that \bghostphysicalop\ indeed obeys $\{Q, b\}={P^2 \over 2}$, as expected. Next, writing the $b$-ghost as a first order differential operator we observe the identity
\eqnn\identityone
$$ \eqalignno{
P^{m}\hat{\bf{A}}_{m} - D_{\alpha}\hat{\bf{W}}_{\alpha} - \half N^{mn}\hat{\bf{F}}_{mn} &= \partial^{m}{\bf{A}}_{m} - d_{\alpha}{\bf{W}}^{\alpha} + {1\over 4}(\lambda\gamma^{mn}\partial_{\lambda}){\bf{F}}_{mn} \ , &  \identityone
}
$$
which is a direct consequence of eqns. \solutionphyisone-\solutionphyifour. Then, eqns. \actionsofphysicalopone-\actionsofphysicalopfour\ allow us to show that
\eqnn \spbghost
$$ \eqalignno{
\{b , \Psi^{(1)}\} &= P^{m}A_{m} - D_{\alpha}W^{\alpha} - \half N^{mn}F_{mn} +  Q\Sigma \ , & \spbghost
}
$$
where $\Sigma$ is given by
\eqnn\sigmabghost
$$ \eqalignno{
\Sigma &= P^{m}\sigma_{m} - d_{\alpha}\rho^{\alpha} + (\lambda\gamma^{mn}w)s_{mn} - (\lambda w)s  \ , & \sigmabghost
}
$$
which is consistent with the particle-limit of the stringy relation $\{b, V\} = U$, where $U$ is the usual ghost number zero vertex operator.

\newsec Amplitudes In Pure Spinor Superspace

\seclab\secthree

\noindent
In this section we review Feynman rules in pure spinor superspace, discussing the gauge-fixing needed to invert the propagator. We then show that these Feynman rules are equivalent to Berends-Giele currents and use the latter to compute tree-level scattering amplitudes. Since Feynman rules are derived from an action, they are manifestly crossing-symmetric, while Berends-Giele currents follow from equations of motion and permit modifications that may break manifest crossing symmetry.

\subsec Feynman Rules

\subseclab\secthreeone

\noindent
 Since the pure spinor superfield $\Psi$ in \nonabelian\ contains both fields and antifields in its definition, we have to employ a non-standard gauge-fixing procedure. Using inspiration from string field theory, the so-called Siegel gauge $b_{0}\Psi = 0$ has been proposed \refs{\Cederwalloverview, \Cederwallsupergeometry} as a candidate for such a purpose, where $b_{0}$ can be viewed as the second order differential operator obtained from \bghostphysicalop\ after using the correspondence principle \replacements. In this section, we will be less restrictive and require that $b_{0}\Psi = Q\Xi$ instead, where $\Xi$ is a ghost number -1 superfield depending on non-minimal variables. As we will see, such a weaker gauge fixing is enough for defining propagators. The gauge-fixed action \nonabelian\ then takes the form
\eqnn \nonabelianfixed
$$ \eqalignno{
S &= \int\,d^{10}xd^{16}\theta [dZ]{\cal{N}}\, Tr\bigg[\half \Psi Q\Psi + {g \over 3}\Psi\Psi\Psi + e(b_{0}\Psi - Q\Xi)\bigg]  \ , &\nonabelianfixed
}
$$
where $e$ is a Lagrange multiplier enforcing the gauge condition. The off-shell fields are Lie-algebra valued, defined as $\Psi = \Psi^a T^a$ where $T^a$ are the generators of the color group. We take these generators to satisfy $[T^{a}, T^{b}] = f^{abc}T^{c}$, where $f^{abc}$ are the totally antisymmetric structure constants and the generators are normalized such that ${\rm Tr}(T^a T^b) = \delta^{ab}$.
Note that the usual unintegrated vertex operator $\Psi^{(1)}(x,\theta,\lambda) = \lambda^{\alpha}A_{\alpha}(x,\theta)$, satisfies the weaker gauge condition
\eqnn \bzeropsi
$$ \eqalignno{
b_{0}\Psi^{(1)} &= Q\bigg[\partial^{m}\sigma_{m} - D_{\alpha}\rho^{\alpha} - (\lambda\gamma^{mn}\partial_{\lambda})s_{mn} + 7(\lambda\partial_{\lambda})s\bigg]  \ , & \bzeropsi
}
$$
as shown in Appendix A.1. For the purpose of obtaining the color-dressed Feynman rules, \nonabelianfixed\ can be written as
\eqnn \nonabelianfixedtwo
$$ \eqalignno{
S &= \int\,d^{10}xd^{16}\theta [dZ]{\cal{N}}\, \bigg[ {1\over 2}\Psi^{a} Q\Psi^{a} + {g \over 6}f^{abc}\Psi^{a}\Psi^{b}\Psi^{c} + e^{a}(b_{0}\Psi^{a} - Q\Xi^{a})\bigg] \ , &\nonabelianfixedtwo
}
$$
by evaluating the trace. 
The Feynman rules following from \nonabelianfixedtwo\ are then given by

\medskip
\noindent
{\it{The Propagator}:} Using the relation $\{Q, b_{0}\} = \lform$ on $\Psi^{a}$, which satisfies $b_{0}\Psi^{a} = Q\Xi^{a}$, one gets that $b_{0}Q = \lform$, and so the propagator is defined by
\eqnn \prop
$$ \eqalignno{
{\cal G}^{ab}(Z,Z') &= \delta^{ab}{2 b_{0}\over \lform} \delta (Z-Z') \ . & \prop
}
$$

\medskip
\noindent
 {\it{The 3-point vertex}:} The Chern-Simons like term in \nonabelianfixedtwo\ gives rises to the 3-point vertex
\eqnn \threepoint
$$ \eqalignno{
{\cal V}^{abc} &= g f^{abc} \int d^{10}x d^{16}\theta [dZ]{\cal N} \ . &\threepoint
}
$$

\noindent
The full amplitude is then obtained by a sum over all Feynman graphs, 
\eqnn\amplitude
$$ \eqalignno{
{\cal A }_n &=\sum_{i\in \Gamma_n}  {c_i n_i \over D_i } \ , &
\amplitude
}
$$
where the color-factors $c_i$ are given by contractions of structure constants, the denominators $D_i$ are products of Mandelstam variables, and $n_i$ are the numerators and encode the remaining kinematic dependence, including pure spinor integrations.

The color factors of the amplitudes can be expanded in sums over traces of generator matrices $T^a$, and the full amplitude can then be written as
\eqnn\coamplitude
$$ \eqalignno{
{\cal A }_n &=\sum_{\sigma \in S_{n-1}}  A_{\sigma(1,\ldots,n-1),n}{\rm Tr}(T^{\sigma_1}\ldots T^{\sigma_{n-1}}T^n) \ , & \coamplitude
}
$$
where $S_{k}$ is the set of permutations of $k$ external-state labels, and  the objects $A_{\ldots}$ are called color-ordered amplitudes, each contributing to one trace structure in the full amplitude. They can be obtained directly from the color-ordered Feynman rules which are given by:

\medskip
\noindent
 {\it{The Propagator}:} 
\eqnn \propCO
$$ \eqalignno{
{\cal G}(Z,Z')
 &= {2 b_{0}\over \lform} \delta (Z-Z')  \ . & \propCO
}
$$

\medskip
\noindent
{\it{The color-ordered 3-point vertex}}: It is given by the part of the color-dressed three-point vertex that is proportional to ${\rm Tr}(T^a T^b T^c)$,
\eqnn \threepointCO
$$ \eqalignno{
{\cal V} &= g\int d^{10}x d^{16}\theta [dZ]{\cal N} \ . &\threepointCO
}
$$



\medskip
\noindent
The color-ordered amplitudes are then computed by sums over all planar Feynman diagrams. Throughout this work we focus on the color-ordered amplitudes due to their relative computational simplicity.

\subsec Berends-Giele Currents

\subseclab\secthreetwo

\noindent
The color-ordered amplitudes can equivalently be obtained from Berends-Giele currents \berendsgiele. Starting from eqn. \nonabelianfixed, the equations of motion are
\eqnn \eomfixedone
\eqnn \eomfixedtwo
$$ \eqalignno{
Q\Psi +  \Psi\Psi - g b_{0}(e) &= 0 \ , & \eomfixedone\cr
b_{0}\Psi - Q\Xi &= 0 \ , & \eomfixedtwo 
}
$$
where we have rescaled $\Psi \rightarrow {\Psi \over g}$. Applying $b_{0}$ on both sides of \eomfixedone, and using the nilpotency of the $b$-ghost, one finds that
\eqnn \eomfinal
$$ \eqalignno{
\half \lform\Psi +   b_{0}(\Psi\Psi) &= 0  \ , & \eomfinal
}
$$
where one has to use eqn. \eomfixedtwo, which renders the kinetic term invertible.
We now introduce the $n$-particle perturbiner expansion
 \refs{\selivanov,\Mafrareffive}
\eqnn \multiparticleexp
$$ \eqalignno{
\Psi &= \sum_{P} \Psi_{P}T^{P}e^{k_{P}.x} = \sum_{i}\Psi_{i}T^{a_{i}}e^{k_{i}.x} + \sum_{i,j}\Psi_{ij}T^{a_{i}}T^{a_{j}}e^{k_{ij}.x} + \ldots &\multiparticleexp
}
$$
where $i = 1,\ldots, n$ and $P$ stands for non-empty words in particle labels. Lie algebra generators and momenta carrying multi-indices are defined as $T^{P} = T^{a_{1}} \ldots T^{a_{n}}$, and $k_{P} = k_{1} + \ldots + k_{n}$. Mandelstam variables are defined by $s_P=\half k_P^2$. After plugging \multiparticleexp\ into the equation of motion \eomfinal, and collecting all terms with the same Lie generator products, one obtains
\eqnn \recursionrelationsone
$$ \eqalignno{
QV_{i} &= 0 \ , \ \ \ \  \Psi_{P} = -{1 \over s_{P}} \sum_{QR=P}b_{0}(\Psi_{Q}\Psi_{R})  \ , \ \ \ \ b_{0}V_{i} = Q\Xi_{i} \ . & \recursionrelationsone
}
$$
where $\Xi_{i}$ is the single-particle superfield in the perturbiner expansion of $\Xi$, $V_{i} = \Psi_{i}$, for $i=1,\ldots ,n$, and the sum runs over all the possible deconcatenations of $P$ into the non-empty ordered words $Q$, $R$. For example, the deconcatenations of $12345$ are $(1234,5),(123,45),(12,345)$ and $(1,2345)$. The $n$-point color-ordered amplitude then reads
\eqnn \scatteringamplitude
$$ \eqalignno{
{{A}}_{1\ldots n} &= (-1)^{n}\sum_{PQ=(1\ldots n-1)}\langle \Psi_P\Psi_{Q} V_{n}\rangle  \ , & \scatteringamplitude
}
$$
where ${\Psi}_{1\ldots i}$ is given by the recursive formula \recursionrelationsone, see \Mafrareftwo\ for a similar construction. The product $\Psi_{P}\Psi_{Q}$ is just $\Psi_{1...n-1}$ with the outer propagator $b_{0}/\lform$ stripped off. The prescription to remove the last propagator can be seen as the LSZ reduction in pure spinor superspace; the BRST operator is to be applied from the left in order to cancel the external propagator, then the limit of on-shell momentum is taken. The angle brackets $\langle \ldots \rangle$ represent the application of the pure spinor measure studied in section \sectwotwo. Integration over $x$ then imposes the momentum conserving delta function, which we ignore in what follows, while the integration over pure spinors and $\theta$ is responsible for imposing various contractions between on-shell states. 

The Berends-Giele currents precisely reproduce the amplitudes from the color-ordered Feynman rules. This follows from evaluating the pure spinor integrals in the latter until all delta functions from propagators are localized, and then observing that terms in the sum over deconcatenations in \scatteringamplitude\ are in one-to-one correspondence with the planar diagrams Feynman diagrams.

\subsec Examples 

\subseclab\examples

\noindent 
In what follows we compute a few lower-point amplitudes and show that they reproduce those obtained in \refs{\Mafrarefone,\Mafrareftwo,\Mafrarefsixn}.
We start from the three-point amplitude, using eqn. \scatteringamplitude\ we have
\eqnn\threepointamp
$$
\eqalignno{
A_{123} &= \langle V_1 V_2 V_3 \rangle \ , & \threepointamp
}
$$
which is exactly the three-particle amplitude in pure spinor superspace. Notice that we implicitly used the prescription in \scatteringamplitude, and removed the outer propagator from $\Psi_{12}$.

For the four-point amplitude it is convenient to introduce a diagrammatic representation of the currents,
\eqnn \psithree
$$ \eqalignno{
\Psi_{123} 
&= 
\matrix{
    \epsfbox[0 15 30 50]{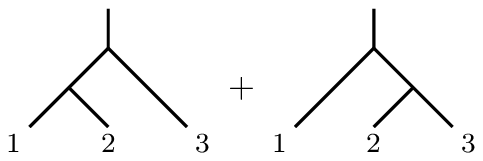}\cr
}
 \cr
&=-{b_0 \over s_{123}}\big(\Psi_{12}V_{3}+V_{1}\Psi_{23}\big) \cr
&={b_0 \over s_{123}}\bigg(
{b_{0}\over s_{12}}\big(V_{1}V_{2}\big)V_{3}+
V_{1}{b_{0}\over s_{23}}\big(V_{2}V_{3} \big) \bigg)  \ , & \psithree
}
$$
where each diagram has a natural Feynman-rule interpretation as compared to those in \Mafratwo.
After removing the last propagator $b_0/s_{123}$ and multiplying with an external state $V_4$, we have
\eqnn \fourpointamplitude
$$
\eqalignno{
A_{1234} &= {1\over s_{12}}\langle b_{0}(V_{1}V_{2})V_{3}V_{4}\rangle + {1\over s_{23}}\langle V_{1}b_{0}(V_{2}V_{3})V_{4}\rangle \ .
& \fourpointamplitude
}
$$
As discussed in A.2 , the action of $b_{0}$ on two single-particle superfields can be demonstrated to take the form 
\eqnn \bghostzerotwo
$$ \eqalignno{
b_{0}(V_{1}V_{2}) &= V_{12} + Q\Lambda_{12} \ , & \bghostzerotwo
}
$$
where $V_{12} = \lambda^{\alpha}A_{12\,\alpha}$, and $A_{12\,\alpha}$ is the 2-particle superfield introduced in \Mafratwo. After dropping BRST-exact terms, eqn. \fourpointamplitude\ becomes
\eqnn \fourpointamplitudefinal
$$
\eqalignno{
{A}_{1234} &= {1 \over s_{12}}\langle V_{12}V_{3}V_{4}\rangle + {1 \over s_{23}}\langle V_{1}V_{23}V_{4}\rangle \ ,
& \fourpointamplitudefinal
}
$$
giving the expected result. 

The five point amplitude can be obtained along similar lines. The four-particle current is given by five planar diagrams,
\eqnn \fourparticlecurrent
$$\eqalignno{
\Psi_{1234} &= 
\matrix{
    \epsfbox[15 10 30 50]{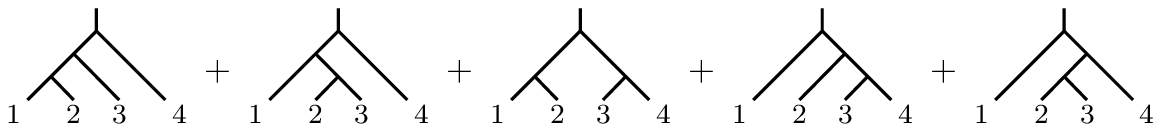}\cr
}
\cr
&= -{b_0\over s_{1234}}
\bigg[ 
{b_0(b_0(V_1 V_2)V_3)V_4 \over s_{123}s_{12}}+
{b_0(V_1 b_0(V_2V_3))V_4 \over s_{123}s_{23}}+
{b_0(V_1 V_2)b_0(V_3 V_4) \over s_{12}s_{34}}\cr
&
\ \ \ \ \ \ \ \ \ \ \ \ \ +
{V_1b_0( V_2 b_0(V_3 V_4)) \over s_{234}s_{34}}+
{V_1b_0( b_0(V_2 V_3) V_4) \over s_{234}s_{23}}
\bigg] \ . & \fourparticlecurrent
}
$$
And from this, the five point amplitude reads
\eqnn \fivepointamplitude
$$ \eqalignno{
A_{12345} &= -{\langle V_{1}b_{0}(V_{2}(b_{0}(V_{3}V_{4}))V_{5}\rangle\over s_{234}s_{34}} - {\langle V_{1}b_{0}(b_{0}(V_{2}V_{3})V_{4})V_{5}\rangle\over s_{234}s_{23}} - {\langle b_{0}(V_{1}V_{2})b_{0}(V_{3}V_{4})V_{5}\rangle\over s_{12}s_{34}}\cr
& - {\langle b_{0}(V_{1}(b_{0}(V_{2}V_{3}))V_{4}V_{5} \rangle \over s_{123}s_{23}} - {\langle b_{0}(b_{0}(V_{1}V_{2})V_{3})V_{4}V_{5} \rangle \over s_{123} s_{12}}  \ . & \fivepointamplitude
}
$$
As shown in A.3, the expressions involving two nested $b_0$ can be calculated to be
\eqnn \twonestedbs
$$ \eqalignno{
b_{0}(b_{0}(V_{i}V_{j})V_{k}) &= V_{ijk} + s_{ij}T_{ijk} + s_{ijk}\Lambda_{ij}V_{k} + Q(\Lambda_{[ij]k}) & \twonestedbs
}
$$
where $V_{ijk} = \lambda^{\alpha}A_{ijk\,\alpha}$ is the 3-particle lowest dimensional superfield studied in \Mafratwo, $T_{ijk}$ is given by
\eqnn \tonetwothree
$$ \eqalignno{
T_{ijk} &= \Lambda_{i}V_{j}V_{k}  - \Lambda_{ij}V_{k} +{\rm cyclic}(ijk) \ , & \tonetwothree
}
$$
and $\Lambda_{[ij]k}$ can be found in 
(A.46).
The amplitude \fivepointamplitude\ then takes the form
\eqnn \fivepointexplicit
$$ \eqalignno{
A_{12345} &= {\langle[V_{342} + s_{34}T_{342} + s_{342}\Lambda_{34}V_{2}]V_{1}V_{5}\rangle \over s_{234} s_{34}} - {\langle[V_{234} + s_{23}T_{234} + s_{123}\Lambda_{23}V_{4}]V_{1}V_{5}\rangle\over s_{234}s_{23}}\cr
& + {\langle (V_{12} + Q\Lambda_{12})(V_{34} + Q\Lambda_{34})V_{5}\rangle \over s_{12}s_{34}}  - {\langle [ V_{231} + s_{23}T_{231} + s_{231}\Lambda_{23}V_{1} ]V_{4}V_{5}\rangle\over s_{123}s_{23}}\cr 
& + {\langle [V_{123} + s_{12}T_{123} + s_{123}\Lambda_{12}V_{3}]V_{4}V_{5}\rangle \over s_{123}s_{12}}
& \fivepointexplicit
}
$$
Cancellations between terms in this expression must happen over common poles, since all objects appearing in the numerator are local. For example, we can focus on terms with the propagator $s_{123}^{-1}$. There are two such contributions
\eqnn\contactcancellingone
$$
\eqalignno{
A_{12345}\bigg|_{s_{123}^{-1}} &= \langle T_{123}V_4 V_5 \rangle - \langle T_{231}V_4 V_5 \rangle \ .
& \contactcancellingone
}
$$
But the $T_{ijk}$ have a cyclic symmetry, and hence these terms do not contribute to the amplitude. Let us also examine the terms with the propagator $s_{34}$, once again there are exactly two such contributions,
\eqnn\contactcancellingtwo
$$\eqalignno{
A_{12345}\bigg|_{s_{34}^{-1}} &= \langle \Lambda_{34}V_2 V_1 V_5 \rangle + {\langle (V_{12} + Q\Lambda_{12})Q\Lambda_{34}V_5 \rangle \over s_{12}} \cr
&=
\langle \Lambda_{34}V_2 V_1 V_5 \rangle
+
{\langle\Lambda_{34}Q (V_{12} + Q\Lambda_{12})V_5 \rangle \over s_{12}}\cr
&=
\langle \Lambda_{34}V_2 V_1 V_5 \rangle
+
\langle\Lambda_{34}V_1 V_2V_5 \rangle \ ,
&\contactcancellingtwo
}
$$
where in the last equality we used that $Q b_0(V_1 V_2)=s_{12}V_1 V_2$, or equivalently that $QV_{12}=s_{12}V_1 V_2$.
A similar analysis follows for the other terms, giving the expected result 
\eqnn \fivepointamplitudefinal
$$ \eqalignno{
A_{12345} &= \langle V_{1}M_{234}V_{5}\rangle + \langle M_{12}M_{34}V_{5}\rangle + \langle M_{123}V_{4}V_{5}\rangle \ , & \fivepointamplitudefinal
}
$$
where
\eqnn \mtwoandthree
$$
\eqalignno{
M_{ij} = {V_{ij}\over s_{ij}} \ ,\ \ \ \ {\rm and} \ \ \ \ 
M_{ijk} &= {1 \over s_{ijk}}\bigg({V_{ijk}\over s_{ij}} - {V_{jki}\over s_{jk}}\bigg) \ . & \mtwoandthree
}
$$
Once the non-minimal variables are decoupled in this way, it is straightforward to evaluate the pure spinor integrals remaining in the expressions \PSS.
In the following section we will argue that our formalism should generate the correct amplitudes to any multiplicity by relating our numerators to those in \Mafraone.

\newsec Properties Of The Numerators

\seclab\secfour

\noindent
The operators used in the previous sections were required to satisfy $b_{0}V = Q\Xi$. We saw in eqn. \bzeropsi\ that the usual unintegrated vertex operator $V = \lambda^{\alpha}A_{\alpha}$, which only depends on minimal variables, satisfies such a gauge condition. In this section we study the properties of the numerators in such a gauge as well as in the Siegel gauge $b_0 V = 0$. Unlike the former, the latter implies that $V$ is a function of non-minimal variables with singularities in $\lambda\bar{\lambda}$. Interestingly though, this gauge will be shown to realize color-kinematics duality, via a mechanism reminiscent of the one in \reiterer. We will restrict our study to the algebraic properties of such a gauge choice, and leave the task of explicit calculations for future work.

\subsec Generalized BRST Blocks And Gauge Invariance

\subseclab\fourone

\noindent
We will first consider the action of the BRST operator on a generic numerator. Each time $Q$ anti-commutes with a $b_0$ we are left with a Mandelstam invariant. In general we have
\eqnn\Qongeneralcurrents
$$
\eqalignno{
Qb_0(b_0(\Psi_\alpha \Psi_\beta)\Psi_\gamma) &= s_{\alpha\beta\gamma}b_0(\Psi_\alpha\Psi_\beta)\Psi_\gamma - s_{\alpha\beta}b_0(\Psi_\alpha\Psi_\beta\Psi_\gamma) + \ldots \ ,
&\Qongeneralcurrents
}
$$
where the Greek letters label some tree diagram, and the ellipses stand for additional terms obtained by propagating the BRST operator through more $b$-ghosts. Focusing on the $s_{\alpha\beta}$ part only, we can represent this diagrammatically,
\eqnn\Qoncurrentsgraphicalone
$$\eqalignno{
&\matrix{
    \epsfbox[0 0 200 50]{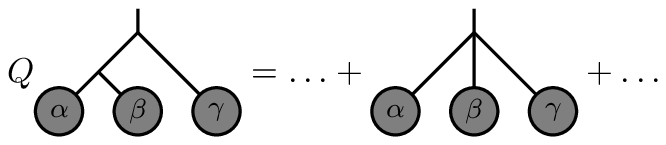}\cr
}
 &\Qoncurrentsgraphicalone
}
$$
where the blobs are place-holders for the unspecified incoming states $\Psi_{\alpha}$, $\Psi_{\beta}$, $\Psi_{\gamma}$. The right-hand-side is proportional to $s_{\alpha\beta}$, hence the corresponding propagator has been contracted in the diagram. In the full amplitude, this term can cancel only with other Feynman diagrams that have the same propagator structure, meaning there must be some other diagram whose BRST variation matches the right-hand-side of the equation above. Indeed, this is
\eqnn\Qoncurrentsgraphicaltwo
$$\eqalignno{
&\matrix{
    \epsfbox[0 0 200 50]{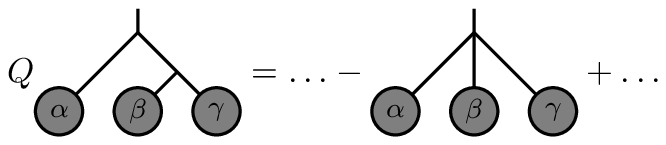}\cr
}
 &\Qoncurrentsgraphicaltwo
}
$$
where the minus sign comes form commuting the BRST operator past the $\alpha$ leg. We notice that diagrams cancel in pairs, where each pair of diagrams differ by exactly one propagator. This observation holds even if we embed our $\alpha\beta\gamma$ sub-diagram in any other larger diagram. 
Let us see an example of this, take the BRST variation of the four-particle half-ladder numerator,
\eqnn \Qonbbbvvvv
$$\eqalignno{
Qb_0(b_0(b_0(V_1 V_2)V_3 )V_4)
&= 
s_{1234}b_0(b_0(V_1 V_2)V_3)V_4 - s_{123}b_0(b_0(V_1 V_2)V_3V_4) \cr
&  \ \ \ + s_{12}b_0(b_0(V_1 V_2 V_3)V_4)
\ . & \Qonbbbvvvv
}
$$
Clearly the $s_{12}$ part has a cyclic symmetry in $(1,2,3)$ which is needed for the cancellation of this term with the $s_{23}$ part coming from the BRST operator acting on $b_0(b_0(V_1b_0( V_2V_3) )V_4)$. Similarly, the term proportional to $s_{123}$ has a cyclic symmetry in $(12,3,4)$, where labels $12$ are treated as a single unit. This is again needed in order to cancel the $s_{34}$ contribution from the variation of $b_0(b_0(V_1 V_2)b_0(V_3 V_4))$.

The transformations of the numerators add up and give the expected equation of motion for Berends-Giele currents \Mafrareftwo
\eqnn \Qonbgcurrent
$$\eqalignno{
Q\Psi_Z &= -\sum_{XY=Z}\Psi_X \Psi_Y \ . & \Qonbgcurrent
}
$$
 without the gauge-fixing terms. 
From this it follows that the amplitudes are BRST-closed, and gauge invariant under transformations of the last leg $\delta V_n = Q\omega$. To see this, consider a gauge transformation of the particle $n$ in an $n-$point amplitude, $V_n\to Q \omega$, it can be written as
\eqnn\variationofamp
$$\eqalignno{
\delta A_{12...n} &= \sum_{XY = (1\ldots n-1)}\langle \Psi_X \Psi_Y Q\omega\rangle \cr
&=
\sum_{XY = (1\ldots n-1)}\langle Q( \Psi_X \Psi_Y) \omega\rangle \cr
&=
\sum_{XY=(1\ldots n-1)}\bigg( 
-\sum_{AB=X}\langle\Psi_A \Psi_B \Psi_Y \omega \rangle+
\sum_{AB=Y}\langle\Psi_X \Psi_A \Psi_B \omega \rangle
\bigg) \cr
&=0 \ . &\variationofamp
}
$$
Notice that a similar argument has been worked out for superstring  amplitudes in \Mafrareffour.
To see that the amplitude is invariant under gauge transformations of any external leg, we first discuss the four-particle numerator as an example.
Under the BRST variation $V_1\to Q\omega$, the same method of anti-commuting the BRST operator with $b$-ghosts gives
\eqnn \Qinbbbvvvv
$$\eqalignno{
b_0(b_0(b_0(Q\omega V_2)V_3 )V_4)
&= 
s_{1234}b_0(b_0(\omega V_2)V_3)V_4 - s_{123}b_0(b_0(\omega V_2)V_3V_4) \cr
&  \ \ \ + s_{12}b_0(b_0(\omega V_2 V_3)V_4)
\ , & \Qinbbbvvvv
}
$$
up to BRST-exact terms.
Once again the $s_{12}$ part above cancels with the $s_{23}$ contribution form $b_0(b_0(V_1b_0( V_2V_3) )V_4)$. The variation of \Qinbbbvvvv\ is identical to \Qonbbbvvvv\ under the replacement $V_1\to \omega$. In this manner, it is not hard to see that under one replacement of $V_i\to Q\omega$ in a numerator, the general transformation rule is
\eqnn\gaugetransform
$$
\eqalignno{
b_0(b_0(...)V_n)\big|_{V_i\to Q\omega} &=\pm \big[Qb_0(b_0(...)V_n)\big]\big|_{V_i\to\omega} \ , &\gaugetransform
}
$$
where the overall sign is determined by the location of the leg being transformed. 
Gauge invariance of the full amplitude can now be shown by substituting this transformation rule into the variation of the amplitude, in a similar fashion to \variationofamp.

Our discussion here mirrors the discussion for the cancellation of contact terms in \twonestedbs, and for good reason; they are both related to gauge invariance. So far we only dealt with linearized gauge invariance of external states, but we actually learned that if we modify our numerators by contact terms that have the same symmetry properties as the ones we discovered in the gauge transformations, they will decouple in the amplitude. For example, one could add to all numerators containing $b_0(b_0(V_i V_j)V_k)$ a term proportional to $s_{ij}$ so long as it is cyclic in $(ijk)$. These kinds of modifications are known as generalized gauge transformations in the amplitudes literature \bcjreview, and reflect the fact that the numerators, which are gauge dependent, are not unique\foot{It is also possible to apply generalized gauge transformations that are polynomial in Mandelstams. These may have cancellations between more than just pairs of diagrams, but they do not play a major role in our analysis.}.

In \refs{\Mafraone,\Mafratwo}, gauge invariance of the amplitudes followed from a multi-particle equation of motion,
\eqnn\nparticleeom
$$\eqalignno{
Q V_{1...n} &= \sum_{j=2}^{n}s_{1...j}\bigg(
\sum_{\alpha\in P(\beta_j)}
V_{1...(j-1)\alpha}V_{j(\beta_j-\alpha)}\cr
& \ \ \ \  \ \ \ \ \ \ \ \ \ \ \ \ \ +
\sum_{\tilde{\alpha}\in P(\beta_{(j+1)})}
V_{1...(j)\tilde{\alpha}}V_{j+1(\beta_{(j-1)}-\tilde\alpha)}
\bigg) \ ,  & \nparticleeom
}
$$
where $\beta_j$ is the range $(j+1,...,n)$ and $P(\beta_j)$ is its power set. 
To unpack this a little, we return to the four-particle example, in which case we have 
\eqnn \QonV
$$\eqalignno{
QV_{1234} &= s_{1234}V_{123}V_4 - s_{123}(V_{123}V_{4}-V_{12}V_{34}-V_{124}V_{3}) \cr
&+ s_{12}(V_{1}V_{234}-V_{12}V_{34}+V_{13}V_{24}+V_{14}V_{23}-V_{124}V_{3}+V_{134}V_{2})\ . &\QonV
}
$$
The important observation for us is that the term proportional to $s_{12}$ has a cyclic symmetry in $(1,2,3)$, just as we observed was necessary for gauge invariance. Now the fact that both our numerators and those in \refs{\Mafratwo} compute the same amplitudes implies that they should be related by some generalized gauge transformation.
To prove this statement in the general case, we make use of Appendices A.2-A.3, where we explicitly show that $b_0(V_1 V_2)$ and $b_0(b_0(V_1 V_2)V_3)$ differ from $V_{12}$ and $V_{123}$ by such generalized gauge transformations, and assume that this same pattern holds up to some multiplicity $n$, that is
\eqnn\contactterm
$$
\eqalignno{
b_0(b_0(\ldots)V_n) &= V_{12...n} + {\cal C}_n \ , & \contactterm
}
$$
where ${\cal C}_n$ is a combination of contact terms possessing the right symmetry to decouple from the Berends-Giele currents at multiplicity $n$, or BRST exact terms that decouple irrespective of their symmetry properties. Given that ${\cal C}_n$ decouples at multiplicity $n$ immediately implies it decouples from all higher multiplicities due to the recursive definition of the Berends-Giele currents (wrapping the contact terms by an additional $b_0$ does not affect their cancellations). Therefore we simply have to show that
\eqnn\contacttermrecursion
$$
\eqalignno{
b_0(V_{12...n}V_{n+1}) &= V_{12...n+1} + {\cal C} \ , &\contacttermrecursion
}
$$
where ${\cal C}$ decouples from currents of order $n+1$. This follows from observing that the multi-particle equations of motion of \Mafratwo, for example \nparticleeom, are equivalent to the single particle equations of motion up to contact terms which have the right symmetries to decouple from amplitudes. We have shown in the appendix that $b_0(V_1V_2)=V_{12}+Q\Lambda_{12}$, and so the multi-particle generalization of this simply includes additional contact terms that decouple. 

Under the operation of $Q$, our numerators transform into a sum of cubic diagrams with one quartic vertex, corresponding to a cancelled propagator. Naively this seems different to the transformation rules in eqn. \nparticleeom, but in fact they are related, with ours generalizing \nparticleeom\ by the addition of more Mandelstam variables. We will return to this at the end of the next subsection, after we develop a better understanding of how the $b$-ghost acts on products of fields.

\subsec Color-Kinematics Duality

\subseclab\fourtwo

\noindent
To start with, it is convenient to write down the $b$-ghost in the simplified form
\eqnn \simplifiedbghost
$$ \eqalignno{
b &= \bigg[P^{m} + {(\lambda\gamma^{mn}r)\over 4 (\lambda\bar{\lambda})}{\bf{A}}_{n}\bigg]{\bf{A}}_{m} \ , & \simplifiedbghost
}
$$
where $\bf{A}_{m}$ was defined in \solutionphyistwo. Up to shift-symmetry terms, the expression inside the square brackets is nothing but $-\Delta_{m}$ defined in \deltadef. Eqn. \simplifiedbghost\ can then be compactly written as
\eqnn \simplifiedbghosttwo
$$ \eqalignno{
b &= -\Delta^{m}{\bf{A}}_{m} \ . & \simplifiedbghosttwo
}
$$
In this manner, when $b_{0}$ acts on the general vertices $V_{1}$, $V_{2}$, one gets
\eqnn \bghostaction
$$ \eqalignno{
b_{0}(V_{1}V_{2}) &= (b_{0}V_{1})V_{2} + \hat{\Delta}^{m}V_{1}\hat{\bf{A}}_{m}V_{2} - \hat{\bf{A}}^{m}V_{1}\hat{\Delta}_{m}V_{2} - V_{1}(b_{0}V_{2}) \ , & \bghostaction
}
$$
where, as before, we are using hatted symbols to represent the operator version of the corresponding fields. Similarly, using the Leibniz rule, one can show the action of $b_{0}$ on three general superfields $V_{1}$, $V_{2}$, $V_{3}$ is given by
\eqnn \bghostactionthree
$$ \eqalignno{
b_{0}(V_{1}V_{2}V_{3}) =&\  b_{0}(V_{1}V_{2})V_{3} +  b_{0}(V_{2}V_{3})V_{1}  + b_{0}(V_{3}V_{1})V_{2} \cr
&- (b_{0}V_{1})V_{2}V_{3}  + V_{1}(b_{0}V_{2})V_{3} - V_{1}V_{2}(b_{0}V_{3})\ .
&  \bghostactionthree
}
$$
When the Siegel gauge condition is imposed on external vertex operators, eqns. \bghostaction, \bghostactionthree\ take the simple form
\eqnn \bghostactionsiegel
\eqnn \bghostactionthreesiegel
$$ \eqalignno{
b_{0}({\cal{V}}_{1}{\cal{V}}_{2}) &= \hat{\Delta}^{m}{\cal{V}}_{1}\hat{\bf{A}}_{m}{\cal{V}}_{2} - \hat{\bf{A}}^{m}{\cal{V}}_{1}\hat{\Delta}_{m}{\cal{V}}_{2} & \bghostactionsiegel\cr
b_{0}({\cal{V}}_{1}{\cal{V}}_{2}{\cal{V}}_{3}) &=  b_{0}({\cal{V}}_{1}{\cal{V}}_{2}){\cal{V}}_{3} +  b_{0}({\cal{V}}_{2}{\cal{V}}_{3}){\cal{V}}_{1}  + b_{0}({\cal{V}}_{3}{\cal{V}}_{1}){\cal{V}}_{2} & \bghostactionthreesiegel
}
$$
where we are using calligraphic letters to denote Siegel gauge operators. The definition of the Poisson bracket
\eqnn \poisson
$$ \eqalignno{
\{X, Y\} &= \hat{\Delta}^{m}X\hat{\bf{A}}_{m}Y
 - \hat{\bf{A}}^{m}X\hat{\Delta}_{m}Y  \ , & \poisson
}
$$
allows us to rewrite eqn. \bghostactionsiegel\ in the suggestive form
\eqnn \bghostsiegelpoisson
$$ \eqalignno{
b_{0}({\cal{V}}_{1}{\cal{V}}_{2}) &= \{{\cal{V}}_{1}, {\cal{V}}_{2}\}  \ . & \bghostsiegelpoisson
}
$$
In addition, the application of $b_{0}$ on both sides of \bghostactionthreesiegel\ gives
\eqnn \bzerojacobi
$$ \eqalignno{
b_{0}(b_{0}({\cal{V}}_{1}{\cal{V}}_{2}){\cal{V}}_{3}) + b_{0}(b_{0}({\cal{V}}_{2}{\cal{V}}_{3}){\cal{V}}_{1}) +
b_{0}(b_{0}({\cal{V}}_{3}{\cal{V}}_{1}){\cal{V}}_{2}) &= 0  \ . & \bzerojacobi
}
$$
This cancellation is precisely all that is needed in order to prove that the Siegel gauge numerators obey the color-kinematics duality\foot{Some subtleties might arise after applying the regularization scheme developed in \Berkovitsn. We elaborate more on this in the Discussions section.}. For external states the identity follows from the Siegel gauge, while internal states are always dressed with a $b$-ghost, and due to the fact that $b_0^2 = 0$, they are effectively in the Siegel gauge too. Let us show this by an example, and also explain how integration by parts within the pure spinor measure allows to deal with the case when one of the particles is the root leg of the diagram.
Consider the six-point ladder numerator
\eqnn\sixpointexample
$$\eqalignno{
n_{123456} &= 
&\matrix{
    \epsfbox[0 0 225 60]{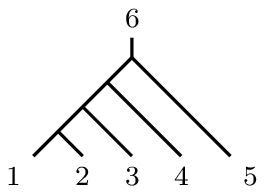}\cr
}
 \cr
 &= 
 \langle b_0(b_0(b_0({\cal{V}}_1 {\cal{V}}_2 ){\cal{V}}_3 ){\cal{V}}_4){\cal{V}}_5 {\cal{V}}_6\rangle \ . &\sixpointexample
}
$$
The Jacobi identity surrounding the $s_{12}$ propagator, that is, the cyclic sum on particles $\{1,2,3\}$ has already been discussed. The Jacobi identity on the $s_{123}$ propagator follows from the cyclic sum on (multi-)particle labels $\{(12),3,4\}$, where $b_0({\cal{V}}_1 {\cal{V}}_2 )$ is treated as a single particle, which is clearly in the Siegel gauge. Finally, we may also take the cyclic sum over the labels $\{4,5,6\}$. It is possible, by use of integration by parts of the $b$-ghost, to show that this is equivalent by relabelling to the Jacobi identity on legs $\{4,5,6\}$. Alternatively, one can use eqn. \bghostactionthreesiegel\ followed by integration by parts to see that this is indeed zero. That integration by parts is possible follows from the fact that $b_{0}$ commutes with the regularization factor ${\cal{N}} = e^{-\lambda\bar{\lambda} - r\theta}$, which can be shown by observing that $[{\bf{A}}_{m}, \lambda\bar{\lambda}] = [{\bf{A}}_{m}, r\theta] = 0$\foot{We thank Carlos Mafra for pointing out a different argument, based on representation theory, which also establishes the validity of this property.}.

The kinematic Jacobi identity is related to the Jacobi identity of the Poisson algebra \poisson , 
\eqnn \bzerojacobipoisson
$$ \eqalignno{
\{\{{\cal{V}}_{1},{\cal{V}}_{2}\},{\cal{V}}_{3}\} + \{\{{\cal{V}}_{2},{\cal{V}}_{3}\},{\cal{V}}_{1}\}  + \{\{{\cal{V}}_{3},{\cal{V}}_{1}\},{\cal{V}}_{2}\} &= 0  \ . & \bzerojacobipoisson
}
$$
That this bracket to obey a Jacobi identity not only on functions in the Siegel gauge it is required that $ [\hat{\Delta}^{m},\hat{\Delta}^{n}]=\{ \hat{\bf{A}}_{m},\hat{\bf{A}}_{m}\}=[\hat{\Delta}^{n},\hat{\bf{A}}_{m}] =0$, which indeed hold. Note that these are also the constraints needed to show that $b_0^2 = 0$. 
This Poisson algebra has a dual Lie algebra with generators
\eqnn \kinematicgenerators
$$\eqalignno{
L_{\psi}:&={\hat \Delta}^m(\psi){\hat{\bf A}}_m - {\hat {\bf A}}_m(\psi) \hat{\Delta}^m \ ,
& 
\kinematicgenerators
}
$$
which obey
\eqnn \kinematicalgebra
$$\eqalignno{
[ L_{\psi},L_{\phi}] &=L_{b_0(\psi\phi)} \ .
& 
\kinematicalgebra
}
$$
For generators defined from functions $\psi$ that satisfy $b_0\psi=0$, this becomes an algebra of infinitesimal diffeomorphisms in pure spinor superspace that preserve the Siegel gauge.

For numerators with particles in the Siegel gauge, there is no difference if one replaces $b_0\to\{\ ,\ \}$. But Siegel gauge external states must be related to any other external states by BRST-exact deformations, ${\cal V}=V+Q(\zeta)$, with $\zeta$ being some superfield depending on non-minimal variables. Since $Q(\zeta)$ is a gauge transformation it decouples from amplitudes, see section \fourone, and we find that the numerators obtained by nested Poisson brackets generate the correct amplitudes with any external states. However, what is lost by this procedure is crossing symmetry, since we cannot integrate by parts the Poisson brackets. Therefore we learn that the Siegel gauge is special in that it allows for both manifest color-kinematics duality and crossing symmetry at the level of individual numerators. 

We have one last observation to make regarding the multi-particle equations of motion. We noted earlier in the section that our numerators obey slightly different equations of motion than \nparticleeom, yet still had all the right transformation properties to ensure that amplitudes are gauge invariant. It can be shown that taking a half-ladder numerator $b_0((...)V_n)$ and operating with $Q$ on the left would reproduce the equation of motion \nparticleeom\ if the external states are in the Siegel gauge. For this, one needs to use the identity \bzerojacobi, and identify $V_{i...jk}$ with the nested $b$-ghost numerators, $b_0((...){\cal V}_k)$. So if external states are not in the Siegel gauge we find that \nparticleeom\ is contaminated by terms with $b_0(V_i)$ which contribute more Mandelstam variables. These generalize eqn. \nparticleeom\ by adding to the right-hand side terms that are polynomial in Mandelstams and are functions of $\Lambda_i$ as well as $V_i$.

\newsec Discussions and Future Research Directions 

\seclab\secfive

\noindent
In this work we studied amplitudes obtained directly from the 10D pure spinor action for super-Yang-Mills theory. We employed Berends-Giele currents, which streamline computations for tree-level amplitudes, and showed that they are equivalent to the amplitudes obtained from the particle-limit of pure spinor open superstrings. We have made extensive use of physical operators, finding that they dramatically simplify computations. In the study of properties of individual numerators we found that the color-kinematics duality emerged from the second-order Leibniz rule obeyed by the $b$-ghost, which is the propagator-numerator for our Feynman rules. Therefore color-kinematics duality emerges directly from the Feynman rules. This is in contrast to other constructions where no mention of an action is made \cheung, or where the color-kinematics duality is enforced by taking gauge-transformations of the kinematic numerators \refs{\Mafratwo,\Mafrathree,\bridges,\reiterer}.

\medskip
Individual numerators are harder to obtain in this formalism than complete amplitudes, since each numerator is given by a pure spinor superspace integral of nested $b$-ghosts acting on on-shell states, and one cannot decouple the non-minimal variables from it. As discussed in section \sectwotwo, when the numerator's integrand diverges slower than $\lambda^{-8}\bar{\lambda}^{-11}$, the integral is well-defined. Since $b$-ghosts carry poles in $\lambda{\bar\lambda}$, the poles of the numerators grow with multiplicity and eventually exceed $\lambda^{-8}\bar{\lambda}^{-11}$. Fortunately, these divergences do not contribute to amplitudes, meaning they completely decouple. However, to define individual numerators these divergences need to be regularized. It was shown in \Berkovitsn\ 
that in principle the $b$-ghost can be regularized in a BRST-invariant manner, so that $b'(z) = b(z) + \{Q, \chi_{\epsilon}(z)\}$, for some $\chi_{\epsilon}(z)$ which depends on the regularization parameter $\epsilon$. If such a $\chi$ can be shown to satisfy $\chi(z) = \{b_{0}, {\cal{K}}(z)\}$, for some conformal weight zero ${\cal{K}}(z)$, then $b_{0}'(z) = e^{-\{Q,{\cal{K}}\}}b_{0}(z)e^{\{Q,{\cal{K}}\}}$, and thus nilpotency of $b_{0}$ is preserved after regularization.
Such an idea has been used in \Karlsson\ for instance.
This idea has been used in \Karlsson\ for studying the UV behavior of 11D supergravity scattering amplitudes. 
In such a scenario, the color-kinematics duality realized by the Siegel gauge operators is maintained after regularization, even though explicit computations of the numerators might still be a challenging problem.

\medskip
There are many additional avenues for further exploration. Color-kinematics duality followed directly from the second order nature of the $b$-ghost, without kinematic structure constants appearing in the action, so it would be interesting to find other theories with a similar action and propagator-numerator. Since it has been observed that the color-kinematics duality has connections to supersymmetry \marco, exploring pure spinor type formulations of theories in lower dimensions \refs{\Cederwallthreed,\Cederwallfourd,\Cederwallsixd} seems promising. Perhaps in lower dimensions numerators would be more tractable for direct computations. In fact, an analogous discussion has been carried out for Chern-Simons theory \Henrikmaor, which has an action formulation similar to the pure spinor action \nonabelian.

\medskip
On the other hand, it would also be interesting to apply the ideas developed in this work to other pure spinor field theories. For instance, the abelian Born-Infeld action in pure spinor superspace \Cederwallborninfeld\ involves the operator $\Delta_{m}$ defined in eqn. \deltadef, so the perturbiner method implies that tree-level Born-Infeld scattering amplitudes are determined from the knowledge of the algebraic properties of the $b$-ghost and $\Delta_{m}$, and their respective actions on single-particle superfields $V$. Likewise, the 11D supergravity pure spinor action \Cederwallsugra\ contains 11D physical operators whose actions on single-particle superfields have recently been calculated in \refs{\Berkovitsmax,\max}. The perturbiner method then tells us that tree-level 11D supergravity scattering amplitudes can be directly obtained from the properties satisfied by these operators and the 11D $b$-ghost. We plan to investigate this in the near future.  

\medskip
Finally, the pure spinor Feynman rules of section \secthreeone\ can, in principle, be used for computing loop-level amplitudes. Since loop-level Feynman diagrams involve ghost particles running in the loop, the calculation of loop-level processes seems to require the use of the other ghost sectors of the pure spinor superfield $\Psi$. It would be interesting to understand such a mechanism in a more transparent way, as well as exploring the realization of on-shell techniques in pure spinor superspace, which might dramatically simplify loop-level computations.

\bigskip \noindent{\bf Acknowledgements:} We are grateful to Oliver Schlotterer, Carlos Mafra and Henrik Johansson for providing valuable comments on an earlier version of this work. MG would like to thank Nathan Berkovits for useful discussions. MBS is thankful to Marco Chiodaroli and Henrik Johansson for collaboration on related topics. 
The research of MG is supported by the European Research Council under ERC-STG-804286 UNISCAMP. The research of MBS is supported by the Knut and Alice Wallenberg Foundation under grants KAW 2018.0116 (From Scattering Amplitudes to Gravitational Waves) and KAW2018.0162 (Exploring a Web of Gravitational Theories through Gauge-Theory Methods).

\appendix{A}{The $b$-ghost As A Second-Order Differential Operator}

\subsec The $b$-ghost On A Single Superfield

\subseclab\bononeV

\noindent
The action of the $b$-ghost on the physical sector of the pure spinor superfield $V = \lambda^{\alpha}A_{\alpha}$, displayed in \bzeropsi, can be computed by using eqn. \bghostphysicalop\ after implementing the usual correspondence principle, namely $P_{m} \rightarrow \partial_{m}$, $d_{\alpha} \rightarrow D_{\alpha}$, $w_{\alpha} \rightarrow -\partial_{\lambda^{\alpha}}$. In this manner, one has
\eqnn \bghostsecondorder
$$ \eqalignno{
b_{0}V &= \bigg(\partial^{m}\hat{\bf{A}}_{m} - D_{\alpha}\hat{\bf{W}}^{\alpha} + {1\over 4}(\lambda\gamma^{mn}\partial_{\lambda})\hat{\bf{F}}_{mn}\bigg)V & \bghostsecondorder
}
$$
Using eqns. \actionsofphysicalopone-\actionsofphysicalopfour, eqn. \bghostsecondorder\ becomes
\eqnn \bghostsecondordertwo
$$ \eqalignno{
b_{0}V &= Q\bigg[\partial^{m}\sigma_{m} - D_{\alpha}\rho^{\alpha} - (\lambda\gamma^{mn}\partial_{\lambda})s_{mn}\bigg] - (\lambda D)s + {9\over 4}(\lambda\gamma^{n}g_{n}) + {1 \over 4}\lambda^{\alpha}(\lambda\gamma^{m})_{\beta}\partial_{\lambda^{\alpha}}g_{m}^{\beta}\cr
& - {45 \over 2}(\lambda g) - {5 \over 2} \lambda^{\alpha}\lambda^{\beta}\partial_{\lambda^{\alpha}}g_{\beta}\cr
&= Q\bigg[\partial^{m}\sigma_{m} - D_{\alpha}\rho^{\alpha} - (\lambda\gamma^{mn}\partial_{\lambda})s_{mn}\bigg] - (\lambda D)s + 2(\lambda\gamma^{n}g_{n}) -20 (\lambda g) + {5\over 4}(\lambda D)s - {5\over 4}Qs\cr
& - {(\lambda\gamma^{mn}\bar{\lambda})\over 4(\lambda\bar{\lambda})}(\lambda D)s_{mn} + {(\lambda\gamma^{mn}\bar{\lambda})\over 4(\lambda\bar{\lambda})}Qs_{mn}  &\bghostsecondordertwo
}
$$
where we used the identities
\eqnn \ione
\eqnn \itwo
$$ \eqalignno{
\partial_{\lambda^{\alpha}}g_{m}^{\beta} =& -{ \bar{\lambda}_{\alpha}\over (\lambda\bar{\lambda})}g_{m}^{\beta} + {4(\gamma^{n}\bar{\lambda})^{\beta}\over (\lambda\bar{\lambda})}\bigg[D_{\alpha}s_{mn} - Q({ \bar{\lambda}_{\alpha}\over (\lambda\bar{\lambda})}s_{mn})\bigg] & \ione\cr
\partial_{\lambda^{\alpha}}g_{\beta} =& -{\bar{\lambda}_{\alpha}\over (\lambda\bar{\lambda})}g_{\beta} - {\bar{\lambda}_{\alpha}\over 2(\lambda\bar{\lambda})}\bigg[D_{\beta}s - Q({\bar{\lambda}_{\beta} \over (\lambda\bar{\lambda})}s)\bigg] \cr 
& + {(\gamma^{mn}\bar{\lambda})^{\beta} \over 2(\lambda\bar{\lambda})}\bigg[D_{\beta}s_{mn} - Q({\bar{\lambda}_{\beta}\over (\lambda\bar{\lambda})}s_{mn})\bigg] & \itwo
}
$$
Likewise, the use of eqns. \shiftone\ allows us to state that
\eqnn \lambdags
$$ \eqalignno{
2(\lambda\gamma^{n}g_{n}) - 20(\lambda g) &= {(\lambda\gamma^{mn}\bar{\lambda}) \over 2(\lambda\bar{\lambda})}F_{mn} - 2{(\lambda\gamma^{mn}\bar{\lambda})\over (\lambda\bar{\lambda})}Qs_{mn} + 10Qs &\lambdags
}
$$
Therefore,
\eqnn \bghostsecondorderthree
$$ \eqalignno{
b_{0}V &= Q\bigg[\partial^{m}\sigma_{m} - D_{\alpha}\rho^{\alpha} - (\lambda\gamma^{mn}\partial_{\lambda})s_{mn}\bigg] + {1\over 4}(\lambda D)s - {5\over 4}Qs + {(\lambda\gamma^{mn}\bar{\lambda})\over 2(\lambda\bar{\lambda})}F_{mn} \cr 
& -{(\lambda\gamma^{mn}\bar{\lambda}) \over 4(\lambda\bar{\lambda})}(\lambda D)s_{mn} + {(\lambda\gamma^{mn}\bar{\lambda})\over 4(\lambda\bar{\lambda})}Qs_{mn} - 2{(\lambda\gamma^{mn}\bar{\lambda}) \over (\lambda\bar{\lambda})}Qs_{mn} + 10Qs  & \bghostsecondorderthree
}
$$
This expression can be simplified further by making use of the pure spinor constraints. Indeed, one can show that
\eqnn \qsmn
$$ \eqalignno{
{(\lambda\gamma^{mn}\bar{\lambda}) \over (\lambda\bar{\lambda})}Qs_{mn} &= {\bar{\lambda}Q\xi\over (\lambda\bar{\lambda})} + Qs & \qsmn
}
$$
and so, eqn. \bghostsecondorderthree\ takes the form
\eqnn \bghostsecondorderfour
$$ \eqalignno{
b_{0}V &= Q\bigg[\partial^{m}\sigma_{m} - D_{\alpha}\rho^{\alpha} - (\lambda\gamma^{mn}\partial_{\lambda})s_{mn}\bigg] + {1\over 4}(\lambda D)s - {5\over 4}Qs + 8(\lambda D)s - {5\over 4}(\lambda D)s   \cr 
& - 7(\lambda D)s - {7 \over 4}Qs + 10 Qs\cr 
&= Q\bigg[\partial^{m}\sigma_{m} - D_{\alpha}\rho^{\alpha} - (\lambda\gamma^{mn}\partial_{\lambda})s_{mn} + 7s\bigg] & \bghostsecondorderfour
}
$$
After plugging \shiftone\ into \bghostsecondorderfour, the expression inside the square brackets can be written in the more compact way
\eqnn \morecompactlambda
$$ \eqalignno{
\Lambda &= -{\partial^{m}(\bar{\lambda}\gamma_{m}A)\over (\lambda\bar{\lambda})} + 2{(\bar{\lambda} W)\over (\lambda\bar{\lambda})} + {(\bar{\lambda}\gamma^{m}D) \over 4(\lambda\bar{\lambda})^{2}}(r\gamma_{m}A) & \morecompactlambda
}
$$
As a check, one can easily show this result yields the same expression found by using the Y-formalism $b$-ghost in the minimal pure spinor framework ($r = 0$), defined as $b_{Y} = (Y\gamma^{m}D)\partial_{m}$, with $Y_{\alpha} \rightarrow {\bar{\lambda}_{\alpha} \over (\lambda\bar{\lambda})}$.

\subsec The $b$-ghost On Two Superfields

\subseclab \appendixatwo

\noindent
The action of the $b$-ghost on two single-particle superfields, say $V_{1} = \lambda^{\alpha}A_{1\,\alpha}$, $V_{2} = \lambda^{\alpha}A_{2\,\alpha}$, can be calculated from the following ansatz
\eqnn \bghosttwooperators
$$ \eqalignno{
b_{0}(V_{1}V_{2}) &= V_{12} + Q\Lambda_{12} & \bghosttwooperators
}
$$
where $V_{12} = \lambda^{\alpha}A_{12\,\alpha}$, and $A_{12\,\alpha}$ is given by
\eqnn \aonetwo
$$ \eqalignno{
A_{12\,\alpha} &= {1\over 2}\bigg[(k_{2}\cdot A_{1})A_{2\,\alpha} - (k_{1}\cdot A_{2})A_{1\,\alpha} + (\gamma^{p}W_{1})_{\alpha}A_{2\,p}  - (\gamma^{p}W_{2})_{\alpha}A_{1\,p}\bigg] & \aonetwo
}
$$
which, interestingly, matches the lowest-dimensional multiparticle superfield introduced in \Mafratwo. Notice that the proposal \bghosttwooperators\ is reasonable since both sides of the equality behave in the same way under the action of the BRST charge. Indeed, $Q(b_{0}(V_{1}V_{2})) = s_{12}V_{1}V_{2} = QV_{12}$. To compute $\Lambda_{12}$, one can let the $b$-ghost act on both-sides of \bghosttwooperators\ to get
\eqnn \lambdaonetwo
$$ \eqalignno{
\Lambda_{12} &= -{b_{0}(V_{12})\over s_{12}} & \lambdaonetwo
}
$$
Thus, $\Lambda_{12}$ is defined by eqn. \lambdaonetwo\ up to BRST-exact terms. Let us compute then $b_{0}V_{12}$. For this purpose, it is convenient to first list the equations of motion satisfied by the two-particle superfields 
\eqnn \twoparticleeomone
\eqnn \twoparticleeomtwo
\eqnn \twoparticleeomthree
\eqnn \twoparticleeomfour
$$ \eqalignno{
D_{\alpha}A_{12\,\beta} + D_{\beta}A_{12\,\alpha} &= (\gamma^{m})_{\alpha\beta}A_{12\,m} + (k_{1}\cdot k_{2})(A_{1\,\alpha}A_{2\,\beta} + A_{1\,\beta}A_{2\,\alpha}) &  \twoparticleeomone\cr
D_{\alpha}A_{12\,m} &= (\gamma_{m}W_{12})_{\alpha} + k_{12\,m}A_{12\,\alpha} + (k_{1}\cdot k_{2})(A_{1\,\alpha}A_{2\,m} - A_{2\,\alpha}A_{1\,m}) &  \twoparticleeomtwo\cr
D_{\alpha}W_{12}^{\beta} &= {1\over 4}(\gamma^{mn})_{\alpha}{}^{\beta}F_{12\,mn} + (k_{1}\cdot k_{2})(A_{1\,\alpha}W^{\beta}_{2} - A_{2\,\alpha}W^{\beta}_{1}) &  \twoparticleeomthree\cr 
D_{\alpha}F_{12\,mn} &= k_{12\,m}(\gamma_{n}W_{12})_{\alpha} - k_{12\,n}(\gamma_{m}W_{12})_{\alpha} + (k_{1}\cdot k_{2})\bigg[A_{1\,\alpha}F_{2\,mn} - A_{2\,\alpha}F_{1\,mn}
\cr
& + A_{1\,n}(\gamma_{m}W_{2})_{\alpha} - A_{2\,n}(\gamma_{m}W_{1})_{\alpha} - A_{1\,m}(\gamma_{n}W_{2})_{\alpha} + A_{2\,m}(\gamma_{n}W_{1})_{\alpha}\bigg]\cr 
& & \twoparticleeomfour
}
$$
where,
\eqnn \othermultisuperfieldsone
\eqnn \othermultisuperfieldstwo
\eqnn \othermultisuperfieldsthree
$$ \eqalignno{
A_{12\,m} &= {1\over 2}\bigg[A_{2\,m}(k_{2}\cdot A_{1}) - A_{1\,m}(k_{1}\cdot A_{2}) + (k_{2\,m} - k_{1\,m})(A_{1}\cdot A_{2}) + 2(W_{1}\gamma_{m}W_{2})\bigg]\cr
& & \othermultisuperfieldsone\cr
W_{12}^{\alpha} &= {1\over 4}(\gamma^{mn}W_{2})^{\alpha}F_{1\,mn} + W_{2}^{\alpha}(k_{2}\cdot A_{1}) - {1\over 4}(\gamma^{mn}W_{1})^{\alpha}F_{2\,mn} - W_{1}^{\alpha}(k_{1}\cdot A_{2}) & \othermultisuperfieldstwo\cr
F_{12\,mn} &= k_{12\,m}A_{12\,n} - k_{12\,n}A_{12\,m} - (k_{1}\cdot k_{2})(A_{1\,m}A_{2\,n} - A_{1\,n}A_{2\,m})
& \othermultisuperfieldsthree
}
$$
In this manner, the action of $\bf{A}_{\alpha}$ given in \solutionphyisone\ is readily computed to be
\eqnn \physoponetwoparticle
$$ \eqalignno{
\hat{\bf{A}}_{\alpha}V_{12} &= A_{12\,\alpha} + (\lambda\gamma^{m})_{\alpha}\sigma_{12\,m} & \physoponetwoparticle
}
$$
with,
\eqnn \physoponetwoparticleextra
$$ \eqalignno{
\sigma_{12\,m} &= -{(\bar{\lambda}\gamma_{m}A_{12})\over 2(\lambda\bar{\lambda})} & \physoponetwoparticleextra
}
$$
The action of $\bf{A}_{m}$ in \solutionphyistwo\ can be written as
\eqnn \physoponetwoparticleam
$$ \eqalignno{
\hat{\bf{A}}_{m}V_{12} &= {(\gamma^{m}\bar{\lambda})^{\alpha}\lambda^{\beta}\over 2(\lambda\bar{\lambda})}\bigg[D_{\alpha}A_{12\,\beta} + D_{\beta}A_{12\,\alpha}\bigg] + {(\bar{\lambda}\gamma^{m}\gamma^{p}\lambda)\over 2(\lambda\bar{\lambda})}Q\bigg[-{(\bar{\lambda}\gamma_{p}A_{12})\over 2(\lambda\bar{\lambda})}\bigg] & \physoponetwoparticleam
}
$$
which, after using eqn. \twoparticleeomone, becomes
\eqnn \physoponetwoparticleamtwo
$$ \eqalignno{
\hat{\bf{A}}_{m}V_{12} &= A_{12\,m} - (\lambda\gamma_{m}\rho_{12}) + Q(\sigma_{12\,m}) + (k_{1}\cdot k_{2})\bigg[{1\over 2(\lambda\bar{\lambda})}\bigg(V_{1}(\bar{\lambda}\gamma_{m}A_{2}) - V_{2}(\bar{\lambda}\gamma_{m}A_{1})\bigg)\bigg] \cr 
& & \physoponetwoparticleamtwo
}
$$
where $\rho^{\alpha}_{12}$ is defined by
\eqnn \rhoonetwoalpha
$$ \eqalignno{
\rho_{12}^{\alpha} &= {(\gamma^{p}\bar{\lambda})^{\alpha}\over 2(\lambda\bar{\lambda})}\bigg[A_{12\,p} + Q(\sigma_{12\,p})\bigg] & \rhoonetwoalpha
}
$$ 
Likewise, the action of $\bf{W}^{\alpha}$ in \solutionphyisthree\ takes the form
\eqnn \physoponetwoparticlewalpha
$$ \eqalignno{
\hat{\bf{W}}^{\alpha}V_{12} &= {(\gamma^{m}\bar{\lambda})^{\alpha}\over 2(\lambda\bar{\lambda})}\bigg[-\partial_{m}V_{12} + QA_{12\,m} - (\lambda\gamma_{m}Q\rho_{12})\cr
& + (k_{1}\cdot k_{2})Q\bigg[{1\over 2(\lambda\bar{\lambda})}\bigg(V_{1}(\bar{\lambda}\gamma_{m}A_{2}) - V_{2}(\bar{\lambda}\gamma_{m}A_{1})\bigg)\bigg]\bigg] & \physoponetwoparticlewalpha
}
$$
The use of eqn. \twoparticleeomtwo\ allows us to write \physoponetwoparticlewalpha\ in the form
\eqnn \walphasimplified
$$ \eqalignno{
\hat{\bf{W}}^{\alpha}V_{12} =& \xi_{12}^{\alpha} + (\gamma^{mn}\lambda)^{\alpha}s_{12\,mn} + \lambda^{\alpha}s_{12} + (k_{1}\cdot k_{2})\bigg[{(\gamma^{m}\bar{\lambda})^{\alpha}\over 2(\lambda\bar{\lambda})}\bigg(V_{1}A_{2\,m}- V_{2}A_{1\,m}\bigg)\cr 
& + {(\gamma^{m}\bar{\lambda})^{\alpha}\over 4(\lambda\bar{\lambda})^{2}}\bigg( -V_{1}(r\gamma_{m}A_{2}) + V_{2}(r\gamma_{m}A_{1})\bigg)\bigg] & \walphasimplified
}
$$
where 
\eqnn \shifttermswalpha
$$ \eqalignno{
\xi^{\alpha}_{12} &= W_{12}^{\alpha} - Q\rho_{12}^{\alpha} \ \ \ \ s_{12\,mn} = {(\bar{\lambda}\gamma_{mn}\xi_{12})\over 8(\lambda\bar{\lambda})}\ , \ \ \ s_{12} = {(\bar{\lambda}\xi_{12})\over 4(\lambda\bar{\lambda})} & \shifttermswalpha
}
$$
Furthermore, the action ${\bf{F}}_{mn}$ in \solutionphyifour\ can be calculated to be
\eqnn \fmntwoparticle
$$ \eqalignno{
\hat{\bf{F}}_{mn}V_{12} =& F_{12\,mn} - 4Qs_{12\,mn} - (\lambda\gamma_{[m}g_{12\,n]}) + (\lambda\gamma_{mn}g_{12}) + {(k_{1}\cdot k_{2})\over 2(\lambda\bar{\lambda})}\bigg[V_{1}(\bar{\lambda}\gamma_{mn}W_{2})\cr 
&- V_{2}(\bar{\lambda}\gamma_{mn}W_{1}) +  {(\bar{\lambda}\gamma^{mnp}r)\over 2(\lambda\bar{\lambda})}(V_{1}A_{2\,m} - V_{2}A_{1\,m}) \cr 
& + {(\bar{\lambda}\gamma^{mnp}r)\over 4(\lambda\bar{\lambda})^{2}}\bigg(-V_{1}(r\gamma_{p}A_{2})\} + V_{2}(r\gamma_{m}A_{1})\bigg)\bigg] & \fmntwoparticle
}
$$
where eqn. \twoparticleeomfour\ was used, and
\eqnn \ggtwoparticles
$$ \eqalignno{
r_{12\,mn} = -F_{12\,mn} + 4Qs_{12\,mn} \ &, \ \ \ g_{12\,\alpha} = {(\gamma^{mn}\bar{\lambda})_{\alpha} \over 8(\lambda\bar{\lambda})}r_{12\,mn} - {\bar{\lambda}_{\alpha}\over 2(\lambda\bar{\lambda})}Qs_{12}\cr
g^{\alpha}_{12\,m} &= {(\gamma^{n}\bar{\lambda})^{\alpha}\over (\lambda\bar{\lambda})}r_{12\,mn} & \ggtwoparticles
}
$$
Therefore, one finds that
\eqnn \lambdatwoparticles
$$ \eqalignno{
-\Lambda_{12} &= {b_{0}(V_{12})\over s_{12}}\cr
&= - (A_{1}W^{2} - A_{2}W^{1}) -{2\over (\lambda\bar{\lambda})}\bigg[V_{1}(\bar{\lambda}W_{2}) - V_{2}(\bar{\lambda}W_{1})\bigg] + {k_{12}^{m}\over 2(\lambda\bar{\lambda})}\bigg[V_{1}(\bar{\lambda}\gamma_{m}A_{2}) - V_{2}(\bar{\lambda}\gamma_{m}A_{1})\bigg]\cr
& - {(\bar{\lambda}\gamma^{m}D)\over 2(\lambda\bar{\lambda})}\bigg[V_{1}A_{2\,m} - V_{2}A_{1\,m}\bigg] - {(\bar{\lambda}\gamma^{m}D)\over 4(\lambda\bar{\lambda})^{2}}\bigg[-V_{1}(r\gamma_{m}A_{2}) + V_{2}(r\gamma_{m}A_{1})\bigg]\cr
& + {1\over 4}(\lambda\gamma^{mn}\partial_{\lambda})\bigg[{1\over 2(\lambda\bar{\lambda})}\bigg(V_{1}(\bar{\lambda}\gamma_{mn}W_{2}) - V_{2}(\bar{\lambda}\gamma_{mn}W_{1})\bigg) + {(\bar{\lambda}\gamma^{mnp}r)\over 4(\lambda\bar{\lambda})^{2}}\bigg(V_{1}A_{2\,p} - V_{2}A_{1\,p}\bigg)\cr
& + {(\bar{\lambda}\gamma_{mnp}r)\over 8(\lambda\bar{\lambda})^{3}}\bigg(-V_{1}(r\gamma^{p}A_{2}) + V_{2}(r\gamma^{p}A_{1})\bigg)\bigg]
& \lambdatwoparticles
} 
$$
where we have ignored BRST-exact terms. This expression can be put in a simpler form through the use of pure spinor identities. Indeed, the $r$-independent part of \lambdatwoparticles\ can be cast as
\eqnn \lambdatwoparticlesnor
$$ \eqalignno{
-\Lambda_{12} &= k_{12\,m}V_{1}(\bar{\lambda}\gamma^{m}A_{2}) - k_{12\,m}V_{2}(\bar{\lambda}\gamma^{m}A_{1}) + 2V_{1}(\bar{\lambda}W_{2}) - 2V_{2}(\bar{\lambda}W_{1})\cr
& - {1\over (\lambda\bar{\lambda})}(\lambda\gamma^{mp}\bar{\lambda})A_{1\,m}A_{2\,p} + Q_{0}\bigg[{1\over 2(\lambda\bar{\lambda})}\bigg[(\bar{\lambda}\gamma^{m}A_{1})A_{2\,m} - (\bar{\lambda}\gamma^{m}A_{2})A_{1\,m}\bigg]\bigg]  + O(r)\cr
& & \lambdatwoparticlesnor
}
$$
where we used that
\eqnn \relationone
$$\eqalignno{
-{1\over 2(\lambda\bar{\lambda})}(\bar{\lambda}\gamma^{m}D)\bigg[V_{1}A_{2\,m} - V_{2}A_{1\,m}\bigg] &= Q_{0}\bigg[{1\over 2(\lambda\bar{\lambda})}\bigg[(\bar{\lambda}\gamma^{m}A_{1})A_{2\,m} - (\bar{\lambda}\gamma^{m}A_{2})A_{1\,m}\bigg]\bigg]\cr
& + {k_{12\,m}\over 2(\lambda\bar{\lambda})}(\bar{\lambda}\gamma^{m}A_{1})V_{2} - {k_{12\,m}\over 2(\lambda\bar{\lambda})}(\bar{\lambda}\gamma^{m}A_{2})V_{1} \cr
& + {(\lambda\gamma^{mp}\bar{\lambda})\over (\lambda\bar{\lambda})}A_{1\,p}A_{2\,m} + 5V_{1}(\bar{\lambda}W_{2}) - 5V_{2}(\bar{\lambda}W_{1})\cr
& +{(\bar{\lambda}\gamma^{m}A_{1})\over 2(\lambda\bar{\lambda})}(\lambda\gamma_{m}W_{2}) - {(\bar{\lambda}\gamma^{m}A_{2})\over 2(\lambda\bar{\lambda})}(\lambda\gamma_{m}W_{1}) & \relationone
}
$$
and also,
\eqnn \relationtwo
$$ \eqalignno{
{1\over 4}(\lambda\gamma^{mn}\partial_{\lambda})\bigg[{1\over 2(\lambda\bar{\lambda})}\bigg(V_{1}(\bar{\lambda}\gamma_{mn}W_{2}) -& V_{2}(\bar{\lambda}\gamma_{mn}W_{1})\bigg)\bigg] = {5\over 4(\lambda\bar{\lambda})^{2}}\bigg[V_{2}(\bar{\lambda}W_{1}) - V_{1}(\bar{\lambda}W_{2})\bigg] \cr
& + {1\over 2(\lambda\bar{\lambda})}\bigg[(\lambda\gamma^{m}W_{2})(\bar{\lambda}\gamma_{m}A_{1}) - (\lambda\gamma^{m}W_{1})(\bar{\lambda}\gamma_{m}A_{2})\bigg]\cr
& +{1\over 4(\lambda\bar{\lambda})}\bigg[V_{1}(\bar{\lambda}W_{2}) - V_{2}(\bar{\lambda}W_{1})\bigg] + A_{1}W_{2} - A_{2}W_{1} \cr 
& & \relationtwo
}
$$
On the other hand, the $r$-dependent part can be computed from the following results:
\eqnn \rdependenpartone
$$ \eqalignno{
{(\bar{\lambda}\gamma^{mnp}r)\over 16(\lambda\bar{\lambda})^{2}}\bigg[(\lambda\gamma_{mn}A_{1})A_{2\,p} - (\lambda\gamma_{mn}A_{2})A_{1\,p}\bigg]
&= {1\over 2(\lambda\bar{\lambda})}(r\gamma^{p}A_{1})A_{2\,p} - {(\lambda r)\over 2(\lambda\bar{\lambda})^{2}}(\bar{\lambda}\gamma^{p}A_{1})A_{2\,p}\cr
& - {1\over 4(\lambda\bar{\lambda})^{2}}(\bar{\lambda}\gamma^{m}A_{1})(\lambda\gamma^{p}\gamma_{m}r)A_{2\,p}\cr
& -{1\over 2(\lambda\bar{\lambda})}(r\gamma^{p}A_{2})A_{1\,p} + {(\lambda r)\over 2(\lambda\bar{\lambda})^{2}}(\bar{\lambda}\gamma^{p}A_{2})A_{1\,p}\cr
& + {1\over 4(\lambda\bar{\lambda})^{2}}(\bar{\lambda}\gamma^{m}A_{2})(\lambda\gamma^{p}\gamma_{m}r)A_{1\,p}
& \rdependenpartone
}
$$
\eqnn \rdependenparttwo
$$ \eqalignno{
{(\bar{\lambda}\gamma^{mnp}r)\over 32(\lambda\bar{\lambda})^{2}}\bigg[-&(\lambda\gamma_{mn}A_{1})(r\gamma^{p}A_{2}) + (\lambda\gamma_{mn}A_{2})(r\gamma^{p}A_{1})\bigg] 
= -{(r\gamma^{p}A_{1})\over 4(\lambda\bar{\lambda})^{2}}(r\gamma_{p}A_{2})\cr
& + {(\lambda r)\over 2(\lambda\bar{\lambda})^{3}}(\bar{\lambda}\gamma^{p}A_{1})(r\gamma_{p}A_{2}) - {(\bar{\lambda}\gamma^{m}A_{1})\over 8(\lambda\bar{\lambda})^{3}}(\lambda\gamma_{m}\gamma^{p}r)(r\gamma_{p}A_{2})\cr
& + {(r\gamma^{p}A_{2})\over 4(\lambda\bar{\lambda})^{2}}(r\gamma_{p}A_{1}) - {(\lambda r)\over 2(\lambda\bar{\lambda})^{3}}(\bar{\lambda}\gamma^{p}A_{2})(r\gamma_{p}A_{1})\cr
& + {(\bar{\lambda}\gamma^{m}A_{2})\over 8(\lambda\bar{\lambda})^{3}}(\lambda\gamma_{m}\gamma^{p}r)(r\gamma_{p}A_{1}) & \rdependenparttwo
}
$$
\eqnn \rdependenpartthree
$$ \eqalignno{
{1\over 4(\lambda\bar{\lambda})^{2}}(\bar{\lambda}\gamma^{m}D)\bigg[V_{1}(r\gamma_{m}A_{2}) -& V_{2}(r\gamma_{m}A_{1}) \bigg] = Q_{0}\bigg[-{(\bar{\lambda}\gamma^{m}A_{1})\over 4(\lambda\bar{\lambda})^{2}}(r\gamma_{m}A_{2})+ {(\bar{\lambda}\gamma^{m}A_{2})\over 4(\lambda\bar{\lambda})^{2}}(r\gamma_{m}A_{1})\bigg] \cr
& + {(r\gamma^{m}D)\over 4(\lambda\bar{\lambda})^{2}}\bigg[V_{2}(\bar{\lambda}\gamma_{m}A_{1}) - V_{1}(\bar{\lambda}\gamma_{m}A_{2})\bigg]\cr
& + {1\over 4(\lambda\bar{\lambda})^{2}}\bigg[(\bar{\lambda}\gamma^{m}A_{1})(r\gamma_{m}\gamma^{s}\lambda)A_{2\,s} - (\bar{\lambda}\gamma^{m}A_{2})(r\gamma_{m}\gamma^{s}\lambda)A_{1\,s}\cr
& + (\bar{\lambda}\gamma^{m}\gamma^{s}\lambda)A_{1\,s}(r\gamma_{m}A_{2}) - (\bar{\lambda}\gamma^{m}\gamma^{s}\lambda)A_{2\,s}(r\gamma_{m}A_{1})\bigg] & \rdependenpartthree
}
$$
In this manner, one learns that
\eqnn \lambdaonetwosimplifiedf 
$$ \eqalignno{
\Lambda_{12} &= k_{12\,m}V_{1}(\bar{\lambda}\gamma^{m}A_{2}) - k_{12\,m}V_{2}(\bar{\lambda}\gamma^{m}A_{1}) + 2V_{1}(\bar{\lambda}W_{2}) - 2V_{2}(\bar{\lambda}W_{1}) - {(\lambda\gamma^{mp}\bar{\lambda})\over (\lambda\bar{\lambda})}A_{1\,m}A_{2\,p}\cr
& + {(\lambda\gamma^{p}\gamma_{m}r)\over 4(\lambda\bar{\lambda})^{2}}(\bar{\lambda}\gamma^{m}A_{1})A_{2\,p} - {(\lambda\gamma^{p}\gamma_{m}r)\over 4(\lambda\bar{\lambda})^{2}}(\bar{\lambda}\gamma^{m}A_{2})A_{1\,p} + {(\lambda\gamma_{m}\gamma^{p}r)\over 8(\lambda\bar{\lambda})^{3}}(\bar{\lambda}\gamma^{m}A_{1})(r\gamma_{p}A_{2}) \cr
& - {(\lambda\gamma_{m}\gamma^{p}r)\over 8(\lambda\bar{\lambda})^{3}}(\bar{\lambda}\gamma^{m}A_{2})(r\gamma_{p}A_{1}) + {1\over 4(\lambda\bar{\lambda})^{2}}(r\gamma^{m}D)\bigg[V_{2}(\bar{\lambda}\gamma_{m}A_{1}) - V_{1}(\bar{\lambda}\gamma_{m}A_{2})\bigg]\cr 
& + Q\bigg[{1\over 2(\lambda\bar{\lambda})}\bigg[(\bar{\lambda}\gamma^{m}A_{1})A_{2\,m} - (\bar{\lambda}\gamma^{m}A_{2})A_{1\,m}\bigg]\bigg] & \lambdaonetwosimplifiedf   
}
$$
As before, one can check this expression coincides with that one obtained by using the Y-formalism $b$-ghost in the minimal pure spinor framework ($r=0$), with $Y_{\alpha} \rightarrow {\bar{\lambda}_{\alpha}\over (\lambda\bar{\lambda})}$.

\subsec Two $b$-ghosts On Three Superfields

\subseclab\appendixathree

\noindent
The simplest and non-trival expression involving nested $b$-ghost is that one containing two $b$-ghosts and three superfields, which was relevant for the 5-point amplitude computation \fivepointamplitude. Such an expression reads
\eqnn \nestedtwoap
$$ \eqalignno{
& b_{0}(b_{0}(V_{1}V_{2})V_{3}) & \nestedtwoap
}
$$
One way of computing \nestedtwoap\ is by using the Y-formalism $b$-ghost, which also satisfies $\{Q, b_{Y}\} = \lform$. Then, the $r$-dependent piece is computed by requiring consistency under the action of the BRST charge on both sides of the equality. This computation makes a heavy use of pure spinor and gamma-matrix identities, as well as the 10D super-Yang-Mills equations of motion \tendsym, \twoparticleeomone-\twoparticleeomfour, and thus we will not reproduce it here. An alternative and more elegant way to calculate \nestedtwoap, which can then be generalized to higher-points, is via the 3-particle equation of motion
\eqnn \threeparticleeomzero
$$ \eqalignno{
D_{\alpha}A_{123\,\beta} + D_{\beta}A_{123\,\alpha}  =& (\gamma^{m})_{\alpha\beta}A_{123\,m} + (k_{1}\cdot k_{2})[A_{1\,\alpha}A_{23\,\beta} + A_{2\,\alpha}A_{31\,\beta} + (\alpha \leftrightarrow \beta)]\cr
& + (k_{12}\cdot k_{3})[A_{12\,\alpha}A_{3\,\beta} - (12 \leftrightarrow 3)]
& \threeparticleeomzero
}
$$
where $A_{123\,\alpha}$ is the 3-particle superfield $\hat{A}_{123\,\alpha}$ of \Mafratwo, defined as
\eqnn \athreealpha
$$ \eqalignno{
A_{123\,\alpha} &= \half \bigg[(k_{3}\cdot A_{12} )A_{3\,\alpha} - (k_{12}\cdot A_{3})A_{12\,\alpha} + (\gamma^{p}W_{12})_{\alpha}A_{3\,p} - (\gamma^{p}W_{3})_{\alpha}A_{12\,p}\bigg] & \athreealpha
}
$$
After contracting with pure spinor variables, eqn. \threeparticleeomzero\ becomes
\eqnn \threeparticleeom
$$ \eqalignno{
QV_{123} &= s_{123}V_{12}V_{3} + s_{12}\bigg[V_{1}V_{23} + V_{2}V_{31} + V_{3}V_{12}\bigg] \ ,
& \threeparticleeom
}
$$
Applying $b_{0}$ on both sides of \threeparticleeom\ yields
\eqnn \towardstwonested
$$ \eqalignno{
s_{123}V_{123} - Q(b_{0}(V_{123})) &= s_{123}b_{0}(V_{12}V_{3}) + s_{12}b_{0}\bigg[V_{1}V_{23} + V_{2}V_{31} + V_{3}V_{12}\bigg] & \towardstwonested
}
$$
One can now use eqn. \bghosttwooperators\ to get
\eqnn \towardstwonestedone
$$ \eqalignno{
s_{123}V_{123} - Q(b_{0}(V_{123}))
&= s_{123}\bigg[b_{0}(b_{0}(V_{1}V_{2})V_{3}) + s_{12}\bigg(V_{1}\Lambda_{23} + V_{2}\Lambda_{31} + V_{3}\Lambda_{12}\bigg) - s_{123}\Lambda_{12}V_{3}\bigg]\cr 
& + s_{12}b_{0}(V_{1}b_{0}(V_{2}V_{3})) + s_{12}b_{0}(V_{2}b_{0}(V_{3}V_{1})) + s_{12}b_{0}(V_{3}b_{0}(V_{1}V_{2}))\cr
& + Q\bigg[s_{123} b_{0}(\Lambda_{12}V_{3}) - s_{12}\bigg(b_{0}(V_{1}\Lambda_{23}) + b_{0}(V_{2}\Lambda_{31}) + b_{0}(V_{3}\Lambda_{12})\bigg)\bigg] \cr 
& & \towardstwonestedone
}
$$
The use of the identity then tells us that
\eqnn \towardstwonestedtwo
$$ \eqalignno{
V_{123} &= \bigg[b_{0}(b_{0}(V_{1}V_{2})V_{3}) + s_{12}\bigg(V_{1}\Lambda_{23} + V_{2}\Lambda_{31} + V_{3}\Lambda_{12}\bigg) - s_{123}\Lambda_{12}V_{3}\bigg]\cr 
& - s_{12}\bigg( \Lambda_{1}V_{2}V_{3} + V_{1}\Lambda_{2}V_{3} + V_{1}V_{2}\Lambda_{3}\bigg) + Q\bigg[b_{0}(\Lambda_{12}V_{3}) - {s_{12}\over s_{123}}\bigg[b_{0}\bigg(V_{1}\Lambda_{23}  \cr 
& + V_{2}\Lambda_{31} + V_{3}\Lambda_{12} - \Lambda_{1}V_{2}V_{3} - V_{1}\Lambda_{2}V_{3} - V_{1}V_{2}\Lambda_{3}\bigg)\bigg] + {b_{0}(V_{123})\over s_{123}}\bigg] \cr 
& & \towardstwonestedtwo
}
$$
Therefore,
\eqnn \twonestedbghostsfinal
$$ \eqalignno{
b_{0}(b_{0}(V_{1}V_{2})V_{3}) &= V_{123} + s_{12}T_{123} + s_{123}\Lambda_{12}V_{3} + Q\Lambda_{[12]3} & \twonestedbghostsfinal
}
$$
where $T_{123}$ was defined in \tonetwothree, and $\Lambda_{[12]3}$ is given by
\eqnn \lambdaonetwothree
$$ \eqalignno{
\Lambda_{[12]3} &= -b_{0}\bigg[{s_{12}\over s_{123}}T_{123} + {V_{123}\over s_{123}} + \Lambda_{12}V_{3}\bigg] & \lambdaonetwothree
}
$$
Although this expression appears to not be manifestly local, we have explicitly checked that the $1/s_{123}$ pole cancels in the $r$-independent sector, and $\Lambda_{[12]3}$ just reduces to the multi-particle generalization of $\Lambda_{12}$.

\listrefs

\bye